\documentclass[structabstract]{aa}  
\usepackage[usenames]{color} 

\usepackage{graphicx}
\usepackage{amsmath}	       
\usepackage{multirow}
\usepackage{lscape}

\usepackage{longtable}

\usepackage{./afterpage}

\usepackage{txfonts}
\usepackage{natbib}
\bibpunct{(}{)}{;}{a}{}{,} 

\begin{document}
   \title{The ARAUCARIA project: Grid-Based Quantitative Spectroscopic Study of Massive Blue Stars in \mbox{NGC~55}\thanks{Based on observations obtained at 
  the ESO VLT Large Programme 171.D-0004.}}


   \author{N.~Castro  \inst{1},
				M.~A.~Urbaneja \inst{2},
				A.~Herrero  \inst{3,4},
				M.~Garcia   \inst{3,4},
				S.~Sim\'on-D\'{i}az  \inst{3,4},
				F.~Bresolin  \inst{2},
          			G.~Pietrzy{\'n}ski \inst{5,6},	 
   				R.~-P.~Kudritzki \inst{2,7}
			    \and
				W.~Gieren  \inst{5}				
          }
   \offprints{N. Castro (\email{norberto@noa.gr})}

   \institute{Institute of Astronomy \& Astrophysics, National Observatory of Athens, I. Metaxa \& Vas. Pavlou St., P. Penteli, 15236 Athens, Greece.
   	      \and
                   Institute for Astronomy, 2680 Woodlawn Drive, Honolulu, HI 96822, USA.
	      \and   	   
	           Instituto de Astrof\'{i}sica de Canarias, C/ V\'{i}a L\'{a}ctea s/n, E-38200 La Laguna, Tenerife, Spain.
              \and
	           Dept. de Astrof\'{i}sica, Universidad de La Laguna, Avda. Astrof\'{i}sico Francisco S\'anchez s/n, E-38071 La Laguna, Tenerife, Spain.
              \and
	           Departamento de Astronom\'{i}a, Universidad de Concepci\'on, Casilla 160-C, Concepci\'on, Chile.					   
	      \and
	           Warsaw University Observatory, Al. Ujazdowskie 4,00-478, Warsaw, Poland. 
    	      \and
    	           Max-Planck-Institute for Astrophysics, Karl-Schwarzschild-Str. 1, D-85741 Garching bei M\"unchen, Germany.}


 
  \abstract
   {The quantitative study of the physical properties and chemical abundances of large samples of massive blue stars at different metallicities is a powerful tool to understand the nature and evolution of these objects. Their analysis beyond the Milky Way is challenging, nonetheless it is doable and the best way to investigate their behavior in different environments. Fulfilling this task in an objective way requires the implementation of automatic analysis techniques that can perform the analyses systematically, minimizing at the same time any possible bias.}
   {As part of the ARAUCARIA project we carry out the first quantitative spectroscopic analysis of a sample of 12 B-type supergiants in the galaxy NGC~55 at 1.94\,Mpc away. By applying the methodology developed in this work, we derive their stellar parameters, chemical abundances and provide a characterization of the present-day metallicity of their host galaxy.}
   {Based on the characteristics of the stellar atmosphere/line formation code {\sc fastwind}, we designed and created a grid of models for the analysis of massive blue supergiant stars. Along with this new grid, we implemented a spectral analysis algorithm. Both tools were specially developed to perform fully consistent quantitative spectroscopic analyses of low spectral resolution of B-type supergiants in a fast and objective way.}
   {We present the main characteristics of our {{\sc fastwind}} model grid and perform a number of tests to investigate the reliability of our methodology. The automatic tool is applied afterward to a sample of 12 B-type supergiant stars in NGC~55, deriving the stellar parameters, \ion{Si}{}, \ion{C}{}, \ion{N}{}, \ion{O}{} and \ion{Mg}{} abundances. The results indicate that our stars are part of a young population evolving towards a red supergiant phase. For half of the sample we find a remarkable agreement between spectroscopic and evolutionary masses, whilst for the rest larger discrepancies are present, but still within the uncertainties. The derived chemical composition hints to an average metallicity similar to the one of the Large Magellanic Cloud, with no indication of a spatial trend across the galaxy.}
{The consistency between the observed spectra and our stellar models supports the reliability of our methodology. This objective and fast approach allows us to deal with large samples in an accurate and more statistical way. These are two key issues to achieve an unbiased characterization of the stars and their host galaxies.}

  
    \authorrunning{N. Castro et al.}
    \titlerunning{Quantitative Study of Massive Blue Stars in NGC~55}

   \maketitle

%


\section{Introduction}

The latest generation of large telescopes has opened a wide range of possibilities in the study of massive blue stars, allowing for the first time analyses of resolved stars beyond the Magellanic Clouds, even to nearby galaxies beyond  the limits of our Local Group. This new observational capability is especially important not only to reach a better knowledge of the nature of these objects, but also to understand the chemical and dynamical evolution of their host galaxies (e.g. \citealt{2009ApJ...704.1120U}). The last two decades in particular have witnessed the use of massive stars as reliable metallicity tracers, complementing the classic approach based on \ion{H}{II} regions (see for instance \citealt{2011A&A...526A..48S} and \citealt{2009ApJ...700..309B}), offering at the same time access to  chemical species that are not accessible through \ion{H}{II} region studies (like \ion{Si}{} or \ion{Mg}{}), and at distances where nebular results strongly rely on techniques which need to be carefully calibrated \citep[the so-called strong-line methods, ][]{1979MNRAS.189...95P}. Moreover, blue supergiant stars present themselves as very promising distance indicators, through the application of the wind-momentum--luminosity relationship (WLR, \citealt{1995svlt.conf..246K}) and the flux-weighted gravity--luminosity relationship (FGLR, \citealt{2003ApJ...582L..83K}).

In order to fully understand the nature of these objects, it is required to perform accurate analyses on large samples of massive stars both in the Milky Way and external galaxies. Whilst current multi-object spectrographs are certainly capable of producing such large collections of spectra (e.g. \citealt{2005Msngr.122...36E}), the accurate modeling of their atmospheres is an intrinsically complex task, involving non-local thermodynamical equilibrium processes and strong stellar winds, hence requiring the use of highly sophisticated stellar model atmospheres.  The important advances accomplished in the modeling of massive blue star atmospheres \citep{1997A&A...323..488S,1998ApJ...496..407H,2005A&A...435..669P}, together with the improvement in the computational facilities provide us with the tools to overcome these issues. With regard to the analysis technique, several alternatives have been proposed by different authors to minimize the subjective component present in the widely used {\it by-eye} techniques, by introducing objective, automatic and fast methods, such as: the methodology employed by \citet{2005A&A...441..711M} based on the genetic algorithm PIKAIA \citep{1995ApJS..101..309C}, the grid-based method proposed by \citet{Lefever_2007} or the principal components analysis (PCA) algorithm designed by \citet{2008ApJ...684..118U}. In this work, we present a new automatic grid-based technique implemented over a grid of models calculated with the latest release of the model atmosphere/line formation code {\sc fastwind} \citep{1997A&A...323..488S,2005A&A...435..669P}, specifically designed and optimized for the study of O- and B-type stars in the optical and infrared range. The combination of the {\sc fastwind} code and a grid-based technique enable us to perform, in an objective and fast way, the analysis of B-type supergiants in the optical domain at low spectral resolution. This provides a very efficient way to handle the analysis of data collected by large spectroscopic surveys. 

The growing interest in the nature of massive stars and their host galaxies has propelled the development of several studies within different members of the Local Group, for example: in the nearby Magellanic Clouds \citep{2003A&A...398..455L,2003A&A...400...21R,2004A&A...417..217T,2005ApJ...633..899K,2005A&A...438..265L,2005A&A...434..677T,2007A&A...466..277H,2007A&A...465.1003M,2007A&A...471..625T}, M~31 \citep{2000ApJ...541..610V,2002A&A...395..519T}, M~33 \citep{1998Ap&SS.263..171M,2000ApJ...545..813M,2005ApJ...635..311U,2011ApJ...735...39U}, NGC~6822 \citep{1999A&A...352L..40M,2001ApJ...547..765V}, NGC~3109 \citep{2007ApJ...659.1198E}, WLM \citep{2003AJ....126.1326V,2006ApJ...648.1007B,2008ApJ...684..118U} or IC~1613 \citep{2007ApJ...671.2028B,2010A&A...523A..23G}. Moreover, these studies have  extended to galaxies beyond the Local Group, such as NGC~300  \citep{2002ApJ...567..277B,2005ApJ...622..862U,2008ApJ...681..269K} or NGC~3621 \citep{2001ApJ...548L.159B}, the latter being at a distance of $\sim$6.7\,Mpc.

The ARAUCARIA project\footnote{https://sites.google.com/site/araucariaproject/} (P.I: W. Gieren, \citealt{2005Msngr.121...23G}) is an ambitious project devoted to investigate the effects that the environment could have on different distance indicators. To that end, a number of nearby galaxies (NGC~6822, IC~1613, WLM, NGC~3109, NGC~55, NGC~247, NGC~300 and NGC~7793) have been targeted, both photometrically and spectroscopically. An important part of this ESO long-term project is focused on the young stellar population of these galaxies (see for instance \citealt{2007ApJ...659.1198E} or \citealt{2008ApJ...684..118U}), and one of its main results has been the discovery and partial exploitation of the FGLR of BA supergiant stars as a distance indicator \citep{2003ApJ...582L..83K}. Within the context of the ARAUCARIA project, we presented the first qualitative analysis of massive blue stars in NGC~55 (\citealt{2008A&A...485...41C}, hereafter C08) situated in the Sculptor filament at $1.94\, $Mpc \citep{2006AJ....132.2556P,2008ApJ...672..266G}. We now present the first quantitative analysis on a sample of B-type supergiants in this galaxy. This is a key step for characterizing the prominent population of blue massive stars in NGC~55, noted by \citet{2005Msngr.121...23G}, particularly in terms of their evolutionary status. 

The work presented in this paper is divided in two main parts. In the first one, Sect. \ref{tools}, we present a detailed description of  the tools we have designed for the analysis of massive blue stars at low spectral resolution. The goodness-of-fit criteria and the parameter space covered by our model grid, along with several tests to identify the limitations and the  reliability of our methodology are also presented. In the second part, we analyze 12 early B-type supergiant stars observed in NGC~55, using the previously discussed tools (Sect. \ref{Quantitative}). We will use the results to study the chemical distribution of the present day metallicity of this galaxy, in addition to constraining the evolutionary status of the analyzed massive blue stars, in Sect. \ref{metallicity}. Finally,  Section \ref{Conclusions} provides the final remarks and comments.

\section{A grid-based quantitative analysis technique}
\label{tools}

The quantitative analysis of optical spectra of B-type supergiant stars is based on well established methods (e.g. \citealt{2006A&A...446..279C}, \citealt{2008A&A...481..777S}). At high spectral resolution, the determination of temperature and surface gravity is based on the ionization balance of different ionization stages of the same element (e.g. \ion{Si}{IV}/\ion{Si}{III}), and the fit to Balmer lines wings respectively  (see \citealt{1999A&A...349..553M}). A slightly modified technique is applied in the analysis of low spectral resolution data.  Although the temperature and gravity criteria are the same, restricting the analysis to individual (metal) lines at low spectral resolution is unreliable. The best approach is to reproduce the main features present in the spectrum simultaneously, as it was suggested by \cite{2003ApJ...584L..73U,2005ApJ...622..862U} in the analysis of NGC 300 B-type supergiant stars. This technique has been also successfully applied in the analysis of massive blue stars in WLM \citep{2006ApJ...648.1007B}, NGC~3109 \citep{2007ApJ...659.1198E} and IC~1613 \citep{2007ApJ...671.2028B}.

The complete spectral analysis consists of two steps. In the first one, the fundamental stellar and wind parameters are derived by using a fixed set of models. The determination of the chemical abundances is carried out in a second step by computing tailored models. With the ultimate goal of performing quantitative studies of low resolution spectra of OB-type supergiants at different metallicities in a systematic and objective way, we have implemented an automatic algorithm to determine the stellar parameters by identifying those models in our grid that minimize the differences with respect to the observed optical spectra (a $\chi^2$ minimization). In the next sections we describe the main components of our automatic analysis method.

\subsection{{\sc fastwind} grid of models}
\label{Atmospheric}

The cornerstone of the analysis of massive blue stars is the grid of model atmospheres employed to  reproduce the different features of the spectrum.  Because of its high computational efficiency, we have used the model atmosphere/line formation code {\sc fastwind} \citep{1997A&A...323..488S,2005A&A...435..669P}. This code takes into account NLTE effects in spherical symmetry with an explicit treatment of the stellar wind effects by considering a $\beta$-like wind velocity law \citep{1994A&A...288..231S}, and by ensuring a smooth transition between the pseudo-static photosphere and the inner wind layers. The main advantage with respect to other similar codes is the possibility of generating  realistic models in a short period of time, a crucial point for building large sets of synthetic spectra.


\subsubsection{Stellar parameters}

Each {\sc fastwind} model is described by nine main parameters (we assume that the winds are homogeneous). Ideally, all of them should be considered free in our grid of models. However, this would imply the calculation of a very large number of models to explore the full parameter space. Therefore, we fixed and constrained some of them based on the previous knowledge of the physics of these objects. This saves a significant amount of computing time, without introducing any relevant limitation in the analyses.  A brief description of the criteria used to fix some of the parameters follows, as well as the specific range explored in each case. As a general rule, the boundaries of the model grid were defined aiming at avoiding observed stars too close to the grid's limits.

\begin{itemize}
\item{{\bf{Effective temperature (\boldmath{$T_{\rm{eff}}$}).}} We computed models with temperatures between $9000$ and $35000\,$K,  in steps of $1000\,$K. This interval covers objects with spectral types ranging $\sim$A1\,--\,O8, for solar metallicity.} 

\item{{\bf{Surface gravity (\boldmath{$\log\,g$}).}} Whilst our main interest here is the analysis of supergiant stars, we extended the calculations to higher gravity values. In order to select the gravities  for each temperature, we considered the  FGLR (\citealt{2003ApJ...582L..83K}). As shown by \cite{2008ApJ...681..269K}, this quantity is empirically related to the luminosity of normal blue supergiants 

		\begin{equation}
		M_{\mathrm{bol}}^{\mathrm{FGLR}}=(3.41\pm0.16)\,(\log\,g_{F}-1.5)-(8.02\pm0.04) 
		\label{Eq:log_gf}
		\end{equation}

\noindent where $M_{\mathrm{bol}}^{\mathrm{FGLR}}$ is the bolometric magnitude, and 
$\log\,g_{F}=\log\,g-4\, \log\,(T_{\rm{eff}}\times 10^{-4})$. Supergiant stars of luminosity class Ia and Ib, exhibit  a range of $\log\,g_{F}$ between $\sim1.0$ and $1.5\,$dex. For objects with a temperature of $25000\,$K, this means $\log\,g\sim2.6-3.1\,$dex. Our models were calculated covering a $\log\,g_{F}$ range between $0.9$ and $2.5\,$dex, with an upper limit of $\log\,g=3.7\,$dex (i.e., main sequence stars are not considered in this grid).\\ 

 	\begin{figure}
   		\resizebox{\hsize}{!}{\includegraphics[angle=0,width=\textwidth]{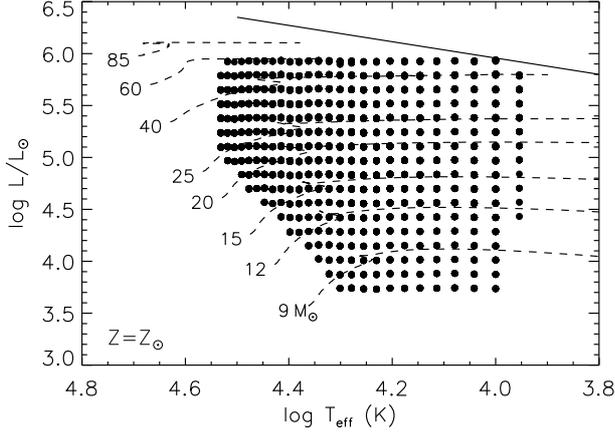}}
   		\caption{Distribution of the models in the $\log\,T_{\rm{eff}}-\log\,L/L_{\odot}$ plane (black dots). To illustrate the stars that can be analyzed with this grid evolutionary tracks with rotation of \cite{2005A&A...429..581M} are plotted. The Humphreys--Davidson Limit is also shown in the upper part of the diagram (solid line). }
    		\label{Fig:red_plot}
 	\end{figure}

Figure \ref{Fig:red_plot} shows the location of our models in the Hertzsprung--Russell (HR) diagram, along with the evolutionary tracks from \cite{2005A&A...429..581M} for an initial equatorial rotation of $300\,$km\,s$^{-1}$ and solar metallicity. As can be deduced from this figure, our parameter selection corresponds to stars with initial masses between $\sim8-60\,M_{\odot}$. Since we have constructed the grid with constant $\log\,g_{F}$ values instead of $\log\,g$, our models define constant luminosity sequences.}

\item{{\bf{Radius (\boldmath{$R_{*}$}).}}  For each given pair $\left[T_{\rm{eff}}, \log\,g\right]$  the radius was calculated from M$_{bol}$ by means of the FGLR (see Eq. \ref{Eq:log_gf}). Note that this relationship was observationally established for supergiant stars, but we have also used it for models that would represent stars that do not belong to this luminosity class. This, however, has no effect on the analysis.}

\item{{\bf{Microturbulence (\boldmath{$\xi$}).}} Three different values of the microturbulence, $7$, $17$ and $27\,\rm{km\,s^{-1}}$, were considered for the calculation of the model atmospheres. A larger number of values were used in the computation of the formal solutions. This procedure does not lead to inconsistencies in the spectrum as long as the value used in the formal solution does not depart far from the value used in the model atmosphere. Hence, the formal solutions were calculated with $\xi$ equal to $5$, $7$, $10$, $12$ km\,$s^{-1}$ (in the case of model atmospheres computed for $7\,\rm{km\,s^{-1}}$), $15$, $17$, $20$, $22$ km\,$s^{-1}$ (for $17\,\rm{km\,s^{-1}}$), $25$, $27$ and $30\,\rm{km\,s^{-1}}$ (for $27\,\rm{km\,s^{-1}}$).}

\item{{\bf{Helium abundance (\ion{He}{}/\ion{H}{}).}} The helium abundance by number is sampled with four points  $0.05$, $0.1$ (solar), $0.15$ and $0.2$. We note here that, although our lowest value is below the primordial He abundance and hence is physically not realistic, it was set to avoid boundary issues, as previously discussed. }

\item{{\bf{Metallicity (Z).}} Five values were used, from  $Z=0.25$ to $1.25\,Z_{\odot}$ in steps of $0.25$, with the solar references taken from \cite{2009ARA&A..47..481A}.  For each metallicity, the elemental abundances (excluding He) are scaled by these values. }

\end{itemize}

	\begin{itemize}
		\item{\textbf{Terminal velocity (\boldmath{$v_{\mathrm{\infty}}$}).} For each model, this parameter was obtained from the escape velocity ($v_\mathrm{esc}$) by using an empirical calibration based on the works by \citet{2000ARA&A..38..613K}, \citet{2006A&A...446..279C} and \citet{2008A&A...478..823M}. According to the studies of \cite{1995ApJ...455..269L}  and  \cite{1999A&A...350..181V} there is a bimodal relationship between both quantities, with an abrupt change at the location of the so-called \textit{'bi-stability jump'}. Contrary to this idea, \cite{2006A&A...446..279C} argued that there is no such break in this relationship. Rather, these authors propose a smooth transition around $20000\,$K. With the goal of emulating the empirical values, a monotonous trend was considered in the range of temperatures where the jump is located,

		\begin{equation}
{\footnotesize
		\frac{v_\mathrm{\infty}}{v_\mathrm{esc}}  = 
	 	 \begin{cases}
	 	  1.10 &  T_{\rm{eff}} \leq 15\,\rm{kK}\\
		  11.65\,\log\,T_{\rm{eff}}-47.62	&  15\,\rm{kK} < T_{\rm{eff}} < 24\,\rm{kK} \\	
	 	  3.41 &  T_{\rm{eff}} \geq 24\,\rm{kK}
	 	 \end{cases}
		\label{Eq:Vinf}
}
		\end{equation}

To account for metallicity effects we assumed that the terminal velocity scales with metallicity as $v_\mathrm{\infty}(Z)\propto Z\,^{0.12}$ (\citealt{1992ApJ...401..596L}, see also \citealt{2000ARA&A..38..613K}, \citealt{2001A&A...369..574V} and \citealt{2002ApJ...577..389K}).
		}

		\item{\textbf{Wind velocity law, \boldmath{$\beta$}.} We adopted an empirical linear relationship between $\beta$ and $T_{\rm{eff}}$, based on results obtained by \citet{2004Miguel_Tesis} and \citet{2006A&A...446..279C} for Galactic B-type supergiants with temperatures in the range $\sim$10000--$\sim$31000 K. Beyond those limits we used fixed values:

		\begin{equation}
{\footnotesize 
		\beta = 
	 	 \begin{cases}
	 	  3.60 &  T_{\rm{eff}} \leq 10\,\rm{kK}\\
	   	  -1.40\, (T_{\rm{eff}} 10^{-4})+5	& 10\,\rm{kK} < T_{\rm{eff}} < 30\,\rm{kK} \\	
  	      0.70 &  T_{\rm{eff}} \geq 30\,\rm{kK}
	 	 \end{cases}
		\label{Eq:beta}
}
		\end{equation}

		}

		\item{\textbf{Mass loss rate  \boldmath{($\dot M$)}.} For each combination of $T_{\rm{eff}}$, $\log\,g$, $\xi$, Z and He/H, we considered 3 different values for the mass loss rate. \cite{1996A&A...305..171P}  showed that different combinations of mass loss rate, terminal velocity and radius can produce the same emergent synthetic profiles as long as the optical depth invariant, defined for smooth winds as $Q = \dot{M} \, / \, \left(R_{*}v_\mathrm{\infty}\right)^{1.5}$, remains constant. Since $R_{*}$ and $v_\mathrm{\infty}$ are not free parameters in our grid, variations of $\dot M$ are equivalent to variations of $Q$. In order to use realistic mass-loss rates for each model, we applied the empirical relationship between the wind momentum and the stellar luminosity found by \citet{1995svlt.conf..246K} and \citet{2000ARA&A..38..613K} to define a proper $Q$-value. The wind momentum--luminosity relationship is defined by
	
\begin{align}
 \log\,D_\mathrm{mom} &\cong x\,\log\,L_{*}/L_{\odot}+D_\mathrm{o} \nonumber  \\
           				    & \cong  2.31 \log\,L_{*}/L_{\odot}+15.94 
\label{Eq:WLR}
\end{align}

\noindent where $D_\mathrm{mom}$ is the modified wind momentum ($\log\,D_\mathrm{mom} = \log\,(\dot{M}v_\mathrm{\infty}R_{*}^{1/2})$). The values for $x$ and $D_\mathrm{o}$ used in Eq. \ref{Eq:WLR} were derived from a linear regression to data from previous studies  \citep{2006A&A...446..279C,2007A&A...473..603M,2007A&A...463.1093L,2008A&A...478..823M}. 

For each set of stellar parameters three different $Q$ values are considered: the one derived from Eq. \ref{Eq:WLR}, $\log\,Q$, and two others with $\log\,Q$ increased/decreased by 0.5\,dex. Finally, we set a lower limit for $\dot{M}$ at $10^{-8}\,M_{\odot}\,\rm{yr}^{-1}$, since at this low mass loss rate the wind effects on the optical profiles are negligible. 
		
The metallicity dependence of the mass-loss rate is accounted for with a power-law, 
$\dot{M}(Z)\propto Z\,^\mathrm{m}$.  For the exponent, we use the results from \cite{2007A&A...473..603M} that found $m=0.83$ based on the analysis of Galactic, LMC and SMC stars.}

	\end{itemize}


\subsubsection{Atomic models}
\label{atomo}

The atomic models used in the calculations will play an important role not only in the determination of stellar parameters but also in the computational time required per model. Detailed atomic models of \ion{H}{I}, \ion{He}{I-II} \citep{Jokuthy_2002}, \ion{Si}{II-III-IV} (N. Przybilla 2007, private communication) and of \ion{O}{II-III} (\citealt{1988A&A...201..232B}; D. Kunze 1998, private communication) are explicitly considered (see below) during the stellar parameters determination. The inclusion of O is critical since at low spectral resolution the line of \ion{Si}{IV} $4116 \AA$ and \ion{Si}{II} $4128-30\ \AA$ are blended with several \ion{O}{II} transitions. During the chemical abundance analysis, detailed models for \ion{C}{II-III} \citep{Eber_1987,1988A&A...202..153E}, \ion{N}{II-III} \citep{Butler_1984,1989A&A...209..244B} and \ion{Mg}{II} (K. Butler 1998, private communication) are also incorporated for the calculation of the tailored models. 

We note here that the other species are treated in an implicit way to account for blanketing/blocking effects. For further details, the reader is referred to \cite{2005A&A...435..669P}.  

\subsection{Stellar parameters determination}
\label{metodo_auto}

\begin{table}
\caption{Spectral features used in the determination of the fundamental parameters.}

\centering
\begin{tabular}{l r | l r }
\hline\hline
Ion & $\lambda\,(\AA)$ & Ion & $\lambda\,(\AA)$ \\
\hline
HI	 & 4101.74     &   HeII	           & 4542.80   \\
HI	 & 4340.47     &   SiII	           & 4128.07	\\ 
HI	 & 4861.33     &   SiII            & 4130.89 \\
HeI	 & 4026.19     &   SiIII	   & 4552.62   \\
HeI	 & 4387.93     &   SiIII	   & 4567.84   \\ 
HeI	 & 4471.48     &   SiIII	   & 4574.76	\\
HeI	 & 4921.93     &    SiIV 	   & 4116.10	                       \\
\hline
\end{tabular}
\tablefoot{At low-resolution the lines of \ion{Si}{II} $4128.07\,\AA$ and \ion{Si}{IV} $4116.10\,\AA$ could be blended with \ion{O}{II} transition.}
\label{Tab:lineasLR}
\end{table}

To avoid a subjective and time-consuming procedure we have implemented a straightforward $\chi^{2}$ technique, similar to the methodology proposed by \cite{2007A&A...463.1093L}. The observed spectrum is compared to a grid of synthetic models, in a number of relevant optical lines. To characterize the goodness-of-fit the differences are evaluated through the following the expression:

		\begin{center}
		\begin{equation}
			\chi^2_\mathrm{i}=\frac{1}{n_\mathrm{lines}}\,\sum_{\mathrm{j=0}}^{n_\mathrm{lines}} \frac{1}{n_{\nu}}\,\sum^{n_{\nu}} \left(\frac {y_\mathrm{ij}-y_\mathrm{obs}}{\sigma}\right) ^{2}
	\label{Eq:chi2}
		\end{equation}
		\end{center}

\noindent where $n_{\nu}$ is the number of wavelength points in the spectral line $j$, $y_\mathrm{obs}$ and $y_\mathrm{ij}$ are the observed and synthetic fluxes respectively (the index $i$ runs over the set of models), and the uncertainty $\sigma$ is estimated according to the signal-to-noise ratio (SNR). Eventually, the average of all the transitions is considered, with all the lines having the same weight. The method was tested giving more weight to those lines that could have more impact on particular stellar parameters (e.g.  silicon transitions in the effective temperature). The results revealed a better match on average when no extra weights were imposed. The selected lines are listed in Table \ref{Tab:lineasLR}. This selection is based on the observed range, the quality of the data and previous knowledge of modeling these spectral features.


 	\begin{figure*}
   		\resizebox{\hsize}{!}{\includegraphics[angle=90,width=\textwidth]{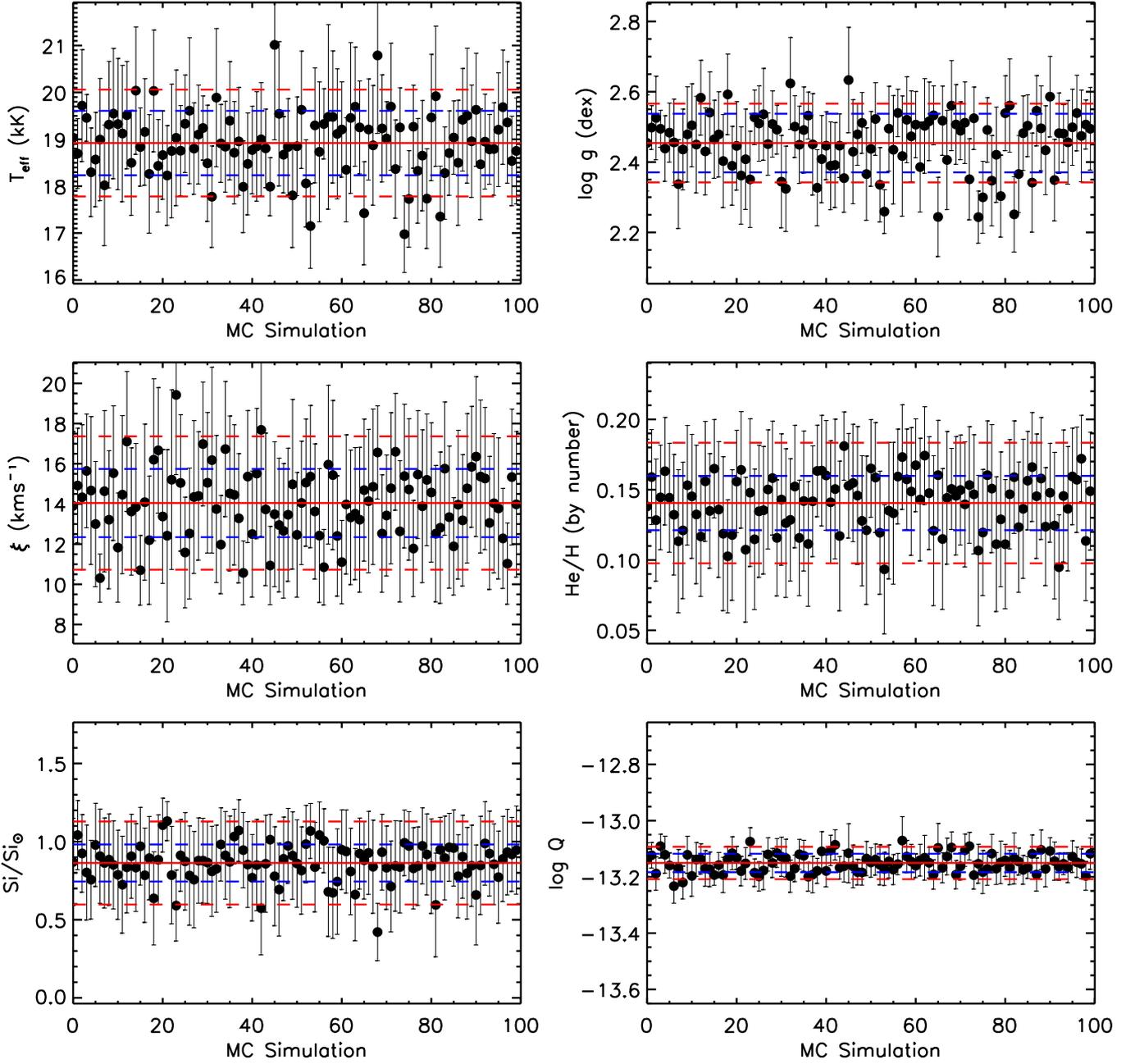}}
 
 		\caption{Stellar parameters derived for the Galactic star HD~14818 ($R=1000$ and $SNR=100$). The panels show the parameters obtained in each Monte Carlo (MC) iteration, the average values are marked by a red solid line (see text for more details). The mean error obtained from the standard deviation in each simulation is marked by red dashed lines while the dispersion in the MC simulation is shown by blue dashed lines.}

     		\label{Fig:MT}
\end{figure*}
\begin{figure*}
   		\resizebox{\hsize}{!}{\includegraphics[angle=90,width=\textwidth]{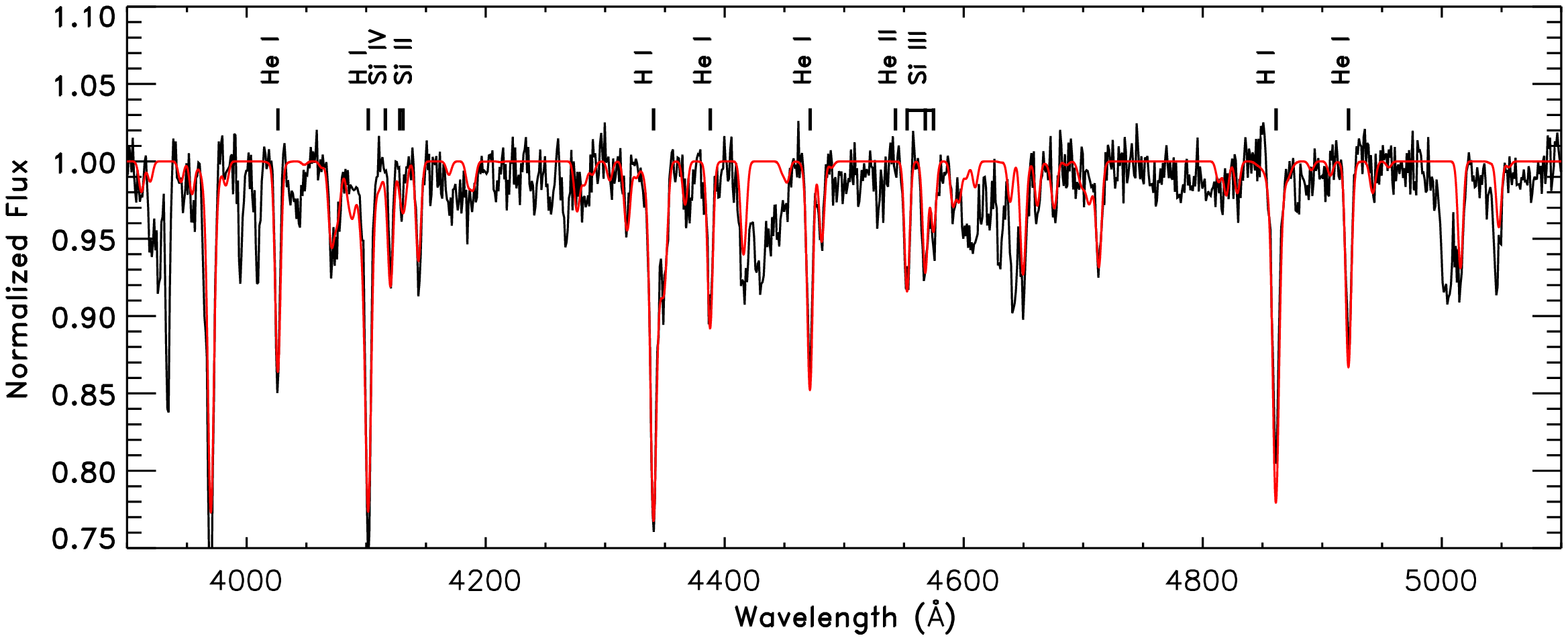}}
   
   		\caption{Best fit model for the spectrum of HD~14818 degraded to $R=1000$ and $SNR=100$. The lines used in the analysis (marked in the plot) are well reproduced. The rest of transitions, some of them interstellar features, were not considered in the analysis.}
     		\label{Fig:hd14818_MT}
 	\end{figure*}


The stellar parameters, and their uncertainties, are determined through two steps, aiming at accounting for different sources of uncertainties:

\begin{itemize}

	\item{From the $\chi^2$ distribution generated by Eq. \ref{Eq:chi2} we selected all those models with a $\chi^2$ value below its $\chi^2$ minimum plus $15\%$. This percentage was not chosen arbitrarily, but using a sample of high resolution spectra with SNR larger than 150 obtained as  part of the IACOB\footnote{The IACOB spectroscopic database  (Sim\'on-D\'{i}az et al., in preparation) presently contains $\sim200$ high quality spectra of O- and B-type Galactic stars ($R=46000$, $SNR>150$)} spectroscopic database. We carried out their analyses by a classic method \citep{1999A&A...349..553M} and with our new $\chi^2$ minimization. By comparing the results, we could identify the percentage that recovered similar errors in both methods. 

	 Those models with a $\chi^2$ that fulfill this criterion were chosen to derive the stellar parameters.  The values and their errors were calculated by averaging all these models, weighted by $e^{-0.5\chi^2}$ (normal distribution probability), and their standard deviations respectively.
	 
	 We investigated the effect that applying different percentage cuts would have on the derived parameters. For example, the difference between a 10\% or 20\%  cut is very small when compared with the errors, with an increment of $\sim200\,$K in the effective temperature uncertainty.}
	
	\item{This method was complemented with a Monte Carlo simulation. Given the SNR of the spectrum, a random array of 100 elements around the continuum position was defined. New re-rectifications of the stellar continuum were performed using these shifts, then the stellar parameters and errors were re-calculated (through the step described before). This allows us to evaluate the impact that the SNR and the continuum rectification have on the results.}
\end{itemize}

Both steps produce a set of errors in the final stellar parameters. We are aware that they are not independent and both estimations are linked to the spectral SNR and the intrinsic uncertainties introduced by the grid design. The final uncertainties result from the quadratic sum of these two uncertainty sources.

\subsubsection{An example: HD~14818}

To illustrate the determination of the stellar parameters, we present the application of our algorithm to the case of the Galactic B-type supergiant HD~14818. A high quality optical spectrum is first degraded to the characteristics of our FORS2 data, $R=1000$ and $SNR=100$. The $\chi^2$ calculations were applied on individual wavelength windows, covering  the spectral features shown in Table \ref{Tab:lineasLR}. Each wavelength range was carefully selected to include the entire line profile. Figure \ref{Fig:MT} displays the results obtained from the Monte Carlo simulation for each individual parameter. As indicated, the spectral resolution, SNR and the set of lines were selected to represent the conditions of the spectra that will be analyzed in the forthcoming section (H$\alpha$ is not included in our analysis since it is not available in the case of the NGC~55 stars).

The quality of the final results is illustrated by Fig. \ref{Fig:hd14818_MT}, where the {\it observed} spectrum of HD~14818 is compared with a model atmosphere computed for the parameters obtained by our automatic analysis algorithm, described in previous sections. Note that only the lines listed in Table \ref{Tab:lineasLR} (marked in the figure) are considered and fitted for the analysis.

\subsubsection{Tests to the method}
\label{test}

Before applying the routine to the analysis of real data, we carried out a number of tests to check the reliability of the proposed methodology. In the first one we analyzed a sample of three synthetic spectra generated with the {\sc fastwind} code. For the second test, we considered the spectra of three Galactic stars (HD~209975 O9.5 Ib, HD~38771 B0.5 Ia and  HD~14818 B2 Ia), originally from the IACOB spectroscopic database. In both cases, the test spectra were degraded to $R=1000$ and $SNR=100$ to simulate our FORS2 NGC~55 spectra. Taking into account that our FORS2 data do not include the  H${\alpha}$ line, we decided to repeat the tests to evaluate the impact of including this line (columns 'Output' and 'Output+H${\alpha}$' in Tables \ref{Tab:Test_Sint}\,--\,\ref{Tab:Test_Real} respectively).

\paragraph{{Comparison to synthetic spectra}} 
\ \\

Three sets of stellar parameters in the range of OB supergiant stars ($\log\,g_{F}\sim 1.0-1.5\,$dex) were randomly chosen from our model grid and analyzed according to the method presented in Sect. \ref{tools}. Table \ref{Tab:Test_Sint} presents the input parameters of the models and the stellar parameters recovered  by the method.  At this spectral resolution, different combinations of parameters, like microturbulence, silicon and  helium abundance, can produce  similar profiles which has a  clear impact on the uncertainties. Nevertheless, our algorithm recovers the input values within the errors.

The analysis yields a better estimate of the wind parameter $Q$ when H$\alpha$ is incorporated. This also produces a slight variation in the rest of parameters, but always within the uncertainties of the previous results obtained without H$\alpha$. It seems possible to constrain $Q$ even without H$\alpha$, just relying on the rest of Balmer lines (mainly H$\beta$). Only for the very hot case ($30900\,$K) we find large differences when H$\alpha$ is not included in the analysis, with a shift of  $0.20\,$dex respect to the input value. For the other two cases, the differences with respect to the input value do not exceed $0.05\,$dex. 

Additional tests were performed with the spectra degraded to $SNR\,=\,50$ (see Table \ref{Tab:Test_Sint}). As could be expected, a wider set of models are compatible with the observations in this case, which translates in larger uncertainties. Nevertheless, the input parameters and the recovered values are in good agreement, always within 1$\sigma$.

\paragraph{{Comparison to Galactic stars}} 
\ \\

The results obtained in the analysis of the three Galactic supergiants are collected in Table \ref{Tab:Test_Real}, along with the parameters derived by \cite{2004A&A...415..349R} and  \cite{2004Miguel_Tesis}. We have restricted the comparison to these two works because we want to minimize the possible effects introduced by the application of different stellar atmosphere codes; here, we are interested in the performance characteristics of our methodology.  Both studies used {\sc fastwind} models, although there are some unavoidable differences. For instance, \cite{2004A&A...415..349R} kept the \newpage

\onecolumn
\begin{landscape}
\begin{table*}
\caption{Results of the analysis of three synthetic spectra calculated with FASTWIND (parameters listed under column \textit{'Input'}) degraded to $R=1000$ and $SNR=100$, $50$. }
{\tiny
\begin{tabular}{l|rrrr|rrrr|rrrr}
\hline\hline
& \multicolumn{4}{c|}{T309G329} & \multicolumn{4}{c|}{T265G280} &\multicolumn{4}{c}{T134G166}  \\ \cline{2-13}
         & Input &  Output & Output+H$\alpha$  & $SNR=50$ &  Input & Output & Output+H$\alpha$ &$SNR=50$ & Input & Output & Output+H$\alpha$  &  $SNR=50$\\
\hline
$T_{\rm{eff}}\,\rm{(kK)}$        &  $30.9$     & $30.8 \pm 0.8$      &   $30.8 \pm 1.1$     &   $30.2 \pm 2.0$        &    $26.5$&     $26.1 \pm 0.8$	       &       $25.8 \pm 1.1$	   &  $25.9 \pm 1.6$    &	$13.4$    &$13.3 \pm 0.6$ &	   $13.2 \pm 0.5$   &  $13.0 \pm 0.7$	\\  
$\log\,g\,\rm{(dex)}$            &  $3.29$     & $3.25 \pm 0.08$     &   $3.26 \pm 0.08$    &   $3.20 \pm 0.15$       &    $2.80$ &    $2.74 \pm 0.11$         &       $2.72 \pm 0.11$     &  $2.74 \pm 0.13$   &	$1.66$    &$1.65 \pm 0.17$ &	   $1.66 \pm 0.07$   & $1.63 \pm 0.16$  \\  
$\xi\,\rm{(km\,s^{-1})}$         &  $14.0$     & $13.6 \pm 3.3$	     &   $14.4 \pm 4.1$     &   $13.0 \pm 6.4$        &    $17.0$ &    $17.5 \pm 4.0$	       &       $17.9 \pm 5.9$	   &  $15.0 \pm 8.0$    &	$20.0$    &$22.5 \pm 5.3$  &	   $19.6 \pm 3.4$    & $20.5 \pm 5.0$	\\  
$He/H$                           &  $0.10$     & $0.12 \pm 0.03$     &   $0.13 \pm 0.04$    &   $0.12 \pm 0.05$       &    $0.12$ &    $0.15 \pm 0.04$         &       $0.14 \pm 0.05$     &  $0.13 \pm 0.05$   &	$0.16$    &$0.15 \pm 0.04$ &	   $0.16 \pm 0.04$   &       $0.12 \pm 0.05$  \\  
$Si/Si_{\odot}$                  &  $1.00$     & $1.01 \pm 0.25$     &   $0.94 \pm 0.24$    &   $0.87 \pm 0.28$       &    $0.75$ &    $0.78 \pm 0.30$         &       $0.66 \pm 0.30$     &  $0.82 \pm 0.0.27$ &	$0.80$    &$0.97 \pm 0.23$ &	   $0.81 \pm 0.23$   &       $0.89 \pm 0.27$  \\  
$\log\,Q  $                      &  $-12.70$   & $-12.90 \pm 0.35$   &   $-12.62 \pm 0.20$  &   $-12.89 \pm 0.33$ 	   & 	$-12.30$  & $-12.30 \pm 0.20$	    & $-12.31 \pm 0.20$   &   $-12.34 \pm 0.20$ &	     $ -12.20$ &$-12.24 \pm 0.20$	&$-12.19 \pm 0.20$&  $-12.28 \pm 0.10$ \\
\hline
\end{tabular}
}
\tablefoot{The resulting parameters without H$\alpha$ are gathered in column \textit{'Output'}. The process was also repeated including H$\alpha$, to estimate the effect of its absence on the stellar parameters (\textit{'Output+H$\alpha$'}). The $SNR=50$ case considers H$\alpha$.}
\label{Tab:Test_Sint}
\caption{Results of the procedure for three Galactic stars degraded to  $R=1000$ and $SNR=100$, $50$. }
{\tiny
\begin{tabular}{l|rrrr|rrrr|rrrr}
\hline\hline
& \multicolumn{4}{c|}{HD 209975} & \multicolumn{4}{c|}{HD 38771} &\multicolumn{4}{c}{HD 14818}  \\ \cline{2-13}
         & R04 & Output & Output+H$\alpha$ & $SNR=50$ &  U04 & Output & Output+H$\alpha$ & $SNR=50$ &  U04   & Output & Output+H$\alpha$ & $SNR=50$\\
\hline
$T_{\rm{eff}}\,\rm{(kK)}$    &$32.0$     & $32.7 \pm 1.0$	 &$32.3 \pm 1.5$    &   $32.0 \pm 3.3$   & $26.5$    &   $27.2 \pm 1.6$   & $26.9 \pm 2.0$	     &   $25.4 \pm 3.3$     &	$20.1$    & $18.3 \pm 0.7$  &	   $18.9 \pm 1.3$	&    $18.9 \pm 1.9$  \\
$\log\,g\,\rm{(dex)}$        &$3.20$     & $3.36 \pm 0.09$ &	 $3.32 \pm 0.08$    &   $3.29 \pm 0.20$  &  $2.90$   &   $3.11 \pm 0.10$  &	  $3.11 \pm 0.11$    &   $3.00 \pm 0.20$    &	$2.40$    & $2.51 \pm 0.11$ &	    $2.45 \pm 0.14$	&    $2.40 \pm 0.18$ \\
$\xi\,\rm{(km\,s^{-1})}$     &$10.0^*$   & $14.7 \pm 5.3$  &	 $14.4 \pm 5.0$     &   $13.9 \pm 7.5$   &  $18.0$   &   $14.5 \pm 4.6$   &	  $14.1 \pm 4.3$     &   $11.3 \pm 6.0$     &	$17.0$    & $15.9 \pm 3.9$  &	    $13.9 \pm 3.7$	&    $11.9 \pm 5.9$  \\
$He/H$                       &$0.10^*$   & $0.11 \pm 0.04$ &	 $0.11 \pm 0.04$    &   $0.12 \pm 0.05$  &  $0.10$   &   $0.15 \pm 0.04$  &	  $0.16 \pm 0.04$    &   $0.14 \pm 0.05$    &	$0.15$    & $0.11 \pm 0.04$ &	    $0.14 \pm 0.05$	&    $0.12 \pm 0.05$ \\
$Si/Si_{\odot}$              &$1.00^*$   & $0.85 \pm 0.30$ &	 $1.08 \pm 0.24$    &   $0.83 \pm 0.28$  &  $1.00^*$ &   $0.91 \pm 0.29$  &	  $0.95 \pm 0.26$    &   $0.81 \pm 0.27$    &	$1.00^*$  & $0.87 \pm 0.30$ &	    $0.87 \pm 0.29$	&    $0.76 \pm 0.27$ \\
$\log\,Q  $                  &$-12.67$   & $-12.69 \pm 0.38$	 &$-12.45 \pm 0.20$ &   $-12.53 \pm 0.25$&  $-12.67$ &   $-13.01 \pm 0.29$&	 $-12.93 \pm 0.25$   &   $-13.11 \pm 0.32$  & $-13.22$    & $-13.03 \pm 0.20$	    &$-13.15 \pm 0.20$   &  $-13.18 \pm 0.11$ \\
\hline
\end{tabular}
}
\tablefoot{The results, summarized in this Table, with the errors (\textit{'Output'}) are compared with analyses performed (though with different observational data) by \cite{2004A&A...415..349R} (\textit{R04}) and \cite{2004Miguel_Tesis} (\textit{U04}). The column \textit{'Output+H$\alpha$'} gives the values returned with the incorporation of H$\alpha$. The $SNR=50$ case considers H$\alpha$. The parameters marked with '*' were kept fixed. }
\label{Tab:Test_Real}
\end{table*}
\end{landscape}

\twocolumn

\noindent microturbulence fixed to $10\,\rm{km\,s^{-1}}$, considering it as a secondary parameter; HD~14818 was analyzed by \cite{2004Miguel_Tesis}  with an early version of {\sc fastwind} that did not include line blocking/blanketing, which explains the difference in temperature (and gravity) with respect to our values.

The main conclusion of these tests is that there is a good agreement between the values obtained by our algorithm applied to the low spectral resolution data and the previous studies based on high spectral resolution data.


 We note that there are important differences in $\log\,Q$ for the three stars, although the values are consistent within the errors. Synthetic H$\alpha$ profiles depend not only on the wind parameters, but also on the effective temperature and surface gravity. Thus, changes in these two fundamental parameters will modify its shape with the consequent adjustment of the value recovered for $Q$. The uncertainties in the stellar parameters will propagate to $\log\,Q$ as well. The differences we have found  could also be a reflection of real changes in the observed profile, due to the use of H$\alpha$ profiles collected in different observational campaigns (H$\alpha$ is variable in these kind of stars, \citealt{2005A&A...440.1133M}). Finally, we should keep in mind that there could be (some) differences inherited from employing different versions of the same model atmosphere code. Nevertheless, in spite of these differences, the values recovered by our method are comparable with the ones suggested by these other works based on high spectral resolution data.

As in the case of the synthetic models discussed above, we performed the analysis of the observed spectra degraded also to $SNR=50$. The outcome of this test is the same; the loss of information due to the lower SNR is reflected in larger uncertainties. 

Both sets of tests confirm the reliability of the technique in finding the main stellar parameters. The quality of the data (SNR) and the available observed wavelength range define the accuracy that can be achieved. We have shown that, by using enough information (see Table \ref{Tab:lineasLR}), it is possible to reach solid results, making our analysis algorithm a very promising tool for the analysis of large collections of optical spectra of OB stars. 

Our main focus is the analysis of supergiant stars. Hence we have designed the grid and selected the spectral features for the analysis accordingly. However, this methodology can be applied to other spectral types/luminosity classes. Of course, this would require different diagnostic transitions, a different set of models, and would present different challenges. The reader is referred to \cite{2010A&A...515A..74L} for a thorough discussion and application of a very similar methodology in the case of Galactic dwarf and giant stars.

 	\begin{figure*}
   		\resizebox{\hsize}{!}{\includegraphics[angle=0,width=\textwidth]{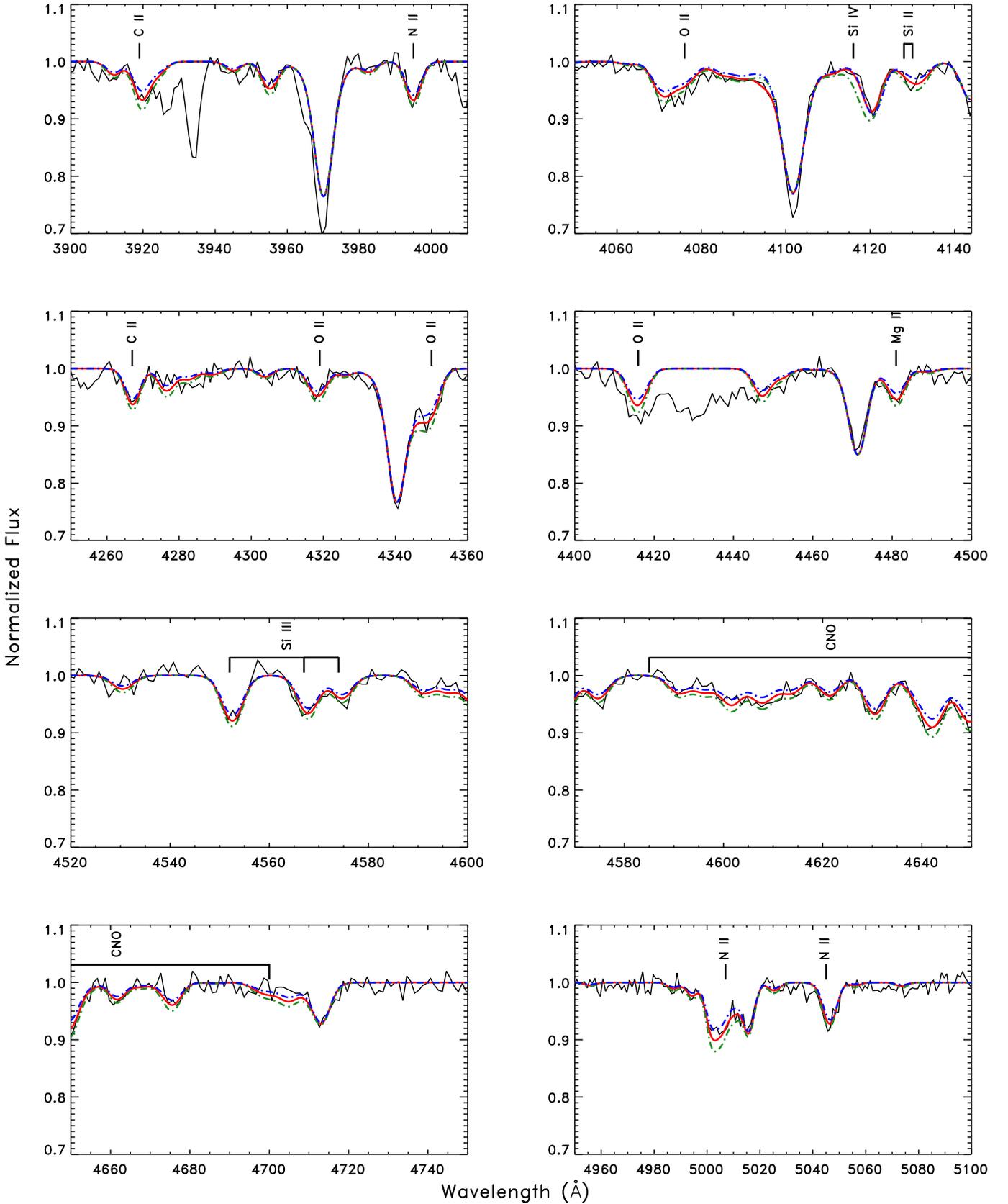}}
 		\caption{Comparison of the final {\sc fastwind} model (red), including the different species considered in the chemical analysis (He, C, N, O, Mg and Si) with the low resolution spectrum of HD~14818 ($R=1000$ and $SNR=100$) in black.  The plot also displays the best model $-0.20$ and $+0.20\,$dex in the abundances of the elements included in  blue and green respectively. The final model provides a good match to the observation. The key diagnostic metal features are identified.}
        \label{Fig:CNO_ejemplo}
	\end{figure*}


\subsection{Abundance determination}
\label{AbunDeter}

Once the fundamental parameters have been determined, we proceed with the analysis of the chemical abundances. As shown by \citet{2003ApJ...584L..73U,2005ApJ...622..862U}, it is possible to derive individual chemical abundances from low resolution optical spectra of early B-type supergiants because of the relative low number density of metal lines (hence minimizing the blends of lines from different species) and because there is a good number of relatively strong features that can be  easily detected. Following these considerations, the methodology applied in the chemical abundance analysis relies on simultaneously modeling all the diagnostic lines for a given species.

Using the derived stellar parameters, a new set of tailored models is computed for each star under analysis by varying the abundances of the relevant species (\ion{C}{}, \ion{N}{}, \ion{O}{} and \ion{Si}{}) in steps of $0.20\,$dex. The chemical analysis is performed in two complementary steps. First, an automatic $\chi^2$ fitting algorithm is used. However, the main features were sometimes weak or blended, and a second visual check is required. 

The abundance uncertainties are estimated from this new set of models, taking into consideration the SNR of the observed spectra. Therefore, the range of abundances defined by the uncertainties, for each individual species, account  for the feature-to-feature scattering, in a similar way as it would be done in a classic analysis.

The precision achieved in the chemical analysis depends not only on the quality of the spectra, but also on the spectral type and on the reliability of the atomic models. For a mid B-type, most of the considered transitions cannot be detected (either they are not present at these temperatures, or they are too weak to be detected at this low resolution and SNR). Note, however, that the \ion{Mg}{II} and \ion{Si}{II} features are stronger for late B-type stars, so we can extract accurate information for those species. In a similar fashion, for a given spectral type and SNR, the abundance uncertainties will depend on the metallicity, with the expectation that  errors become larger with decreasing metallicity, eventually reaching the limit where only upper limits could be placed.

For each individual element, we attended to a particular group of spectral features (see  \citealt{2005ApJ...622..862U} and \citealt{2007ApJ...659.1198E}). Briefly, the lines considered for the chemical analysis,  and summarized in Table \ref{Tab:Lineasabun}, were:

\begin{table*}
\caption{Metal features considered for the chemical analysis.}
\centering
\begin{tabular}{ l r | l r | l r | l r | l r  }
\hline\hline
\multicolumn{2}{c}{Silicon} & \multicolumn{2}{|c|}{Carbon} &\multicolumn{2}{c|}{Nitrogen} &\multicolumn{2}{c|}{Oxygen}  &\multicolumn{2}{c}{Magnesium}\\ \cline{2-10}
\hline
Ion & $\lambda\,(\AA)$ &Ion & $\lambda\,(\AA)$ &Ion & $\lambda\,(\AA)$ &Ion &$\lambda\,(\AA)$ &  Ion & $\lambda\,(\AA)$ \\
\hline
\ion{Si}{IV}     &  $4116$  &   \ion{C}{II}     &  $3919$  &  \ion{N}{II} &  $3995$ & \ion{O}{II}  &  $4076$  & \ion{Mg}{II}  &  $4481$  \\
\ion{Si}{III}    &  $4552$  &   \ion{C}{II}     &  $3921$  &  \ion{N}{II} &  $5007$ & \ion{O}{II}  &  $4319$  & 		       &		 \\   
\ion{Si}{III}    &  $4567$  &	\ion{C}{II}     &  $4267$  &  \ion{N}{II} &  $5045$ & \ion{O}{II}  &  $4350$  & 		       &		 \\   
\ion{Si}{III}    &  $4574$  &                          &                &                      &               & \ion{O}{II}  &  $4416$  &                        &                 \\   
\ion{Si}{II}     &  $4128$  &                          &                &                      &               &  &                        &                 \\   
\ion{Si}{II}     &  $4130$  &                          &                &                      &               &                     &                &                        &                 \\   
\hline
\end{tabular}
\label{Tab:Lineasabun}
\end{table*}

\begin{itemize}

	\item{\textbf{Silicon.} We used the transitions of \ion{Si}{III} $4552-67-74\,\AA$, \ion{Si}{II} $4128-30\,\AA$  and \ion{Si}{IV} $4116\,\AA $. The lines of \ion{Si}{II} are only available at low temperatures, B2 and later spectral types. On the other hand, the profile of \ion{Si}{IV} will be accessible at late O and early B-type stars.  }
	
	\item{\textbf{Oxygen.} The analysis looks for the best fit to the transitions around $4076$, $4319$, $4350$ and $4416\,\AA$; at low spectral resolution they are in fact blends. There are additional \ion{O}{II} features around $4585$ and $4700\,\AA$, but they are blended with other elements. The latter will be used as a consistency check for the abundance derived using the other \ion{O}{II} lines.}
	
	\item{\textbf{Nitrogen.} Some of the most prominent N features in the  wavelength range covered by our FORS2 spectra are blended with other elements (for example around $4650\,\AA$), but there are still some isolated features that can be used to constrain the nitrogen abundance. The study was centered on $3995$, $5007$ and $5045\,\AA$. Note that if nebular subtraction is not accurate, the [\ion{O}{III}] $5007\,\AA$ transition could severely affect the overlapping nitrogen line.}
	
	\item{\textbf{Carbon.} Given the spectral quality and the wavelength range ($\sim3900-5000\,\AA$) the strongest \ion{C}{II} transition is $4267\,\AA$. We consider that the results based (only) on this line are currently not as reliable as for the rest of the species (see \citealt{2003A&A...398..455L}). 
Alternative lines are $3919-21\,\AA$, though too weak in many cases. The \ion{C}{III} transitions around $4650\,\AA$, while blended with O and N, could serve as a secondary check. }
	
	\item{\textbf{Magnesium.} Its abundance is determined using the line of \ion{Mg}{II} $4481\,\AA$.  Note that this transition could be blended with \ion{Al}{III} $4479\,\AA$ \citep{2005MNRAS.358..193L}.}

\end{itemize}

Figure \ref{Fig:CNO_ejemplo} illustrates the results of the chemical analysis of HD~14818. The main diagnostic lines used in the elemental abundance determination, identified in the figure, show a good match with the final model.



\section{Quantitative spectral analysis of NGC~55 B-type supergiant stars}
\label{Quantitative}

In C08 we presented the first spectral catalog of massive blue stars in NGC~55. Very briefly, optical spectra of $\sim$200 sources were collected with the FOcal Reducer/low dispersion Spectrograph 2 (FORS2, \citealt{1998Msngr..94....1A}) at the Very Large Telescope (VLT-UT2). The instrument was equipped with the 600B grism, providing optical spectra in the range $\sim3900-6000\,\AA$ at a resolving power $R \sim1000$. A complete description of the data and the reduction process can be found in the aforementioned reference.

\begin{figure}
   		\resizebox{\hsize}{!}{\includegraphics[angle=0,width=\textwidth]{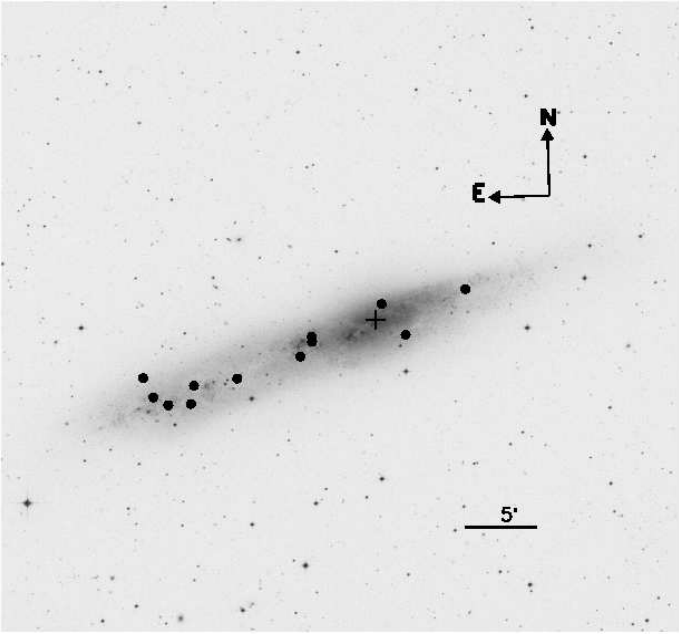}}
   		\caption{Distribution of the objects selected for the analysis over NGC~55. The image was taken 
      from the \textit{DSS} archive  \textit{(http://archive.stsci.edu/cgi-bin/dss\_form)}. The black cross marks the optical galactic center.}
    	\label{Fig:NGC55_cand}
 	\end{figure}

From this previous work, we selected twelve B-type supergiants for a detailed quantitative analysis. The selection was based on the good SNR, spectral type and the lack of obvious contamination by other sources, including strong nebular lines (when possible). The selection of spectral types between late O- and early B-types guarantee that the main spectral features  required for the analysis (see Table \ref{Tab:lineasLR}) are available. The stars are spatially distributed across the galaxy (see Fig. \ref{Fig:NGC55_cand}), which will allow us to investigate the distribution of its chemical composition. Table  \ref{catalog1} summarizes all the relevant information, as well as provides revised photometry. The stars are identified following C08.

\begin{table*}

\caption{NGC~55 stars analyzed in this work.} 

\centering
{\footnotesize
\begin{tabular}{l r r r r r r}
\hline\hline
 ID  &   RA\,(J2000) &   DEC\,(J2000)  & SpT &  SNR & V  &   I    \\
 (1) &  (2) &   (3)  &   (4)  &   (5)  &  (6) &  (7)   \\
\hline
A\_8	&   0:15:44.18	&   -39:14:58.20	&   O9.7I	&   62	&   $20.067\pm0.005$   &  $20.079\pm0.010$ \\
C1\_44	&   0:15:11.29	&   -39:12:34.55	&   B0I		&   79	&   $19.360\pm0.005$   &  $19.331\pm0.007$ \\
C1\_9	&   0:14:45.54	&   -39:12:38.17	&   B1I		&   86	&   $19.304\pm0.004$   &  $19.504\pm0.009$ \\
C1\_13	&   0:14:51.86	&   -39:10:57.35	&   B1I		&   60	&   $19.222\pm0.004$   &  $19.254\pm0.006$ \\
C1\_45	&   0:15:11.31	&   -39:12:50.40	&   B1I		&   38	&   $19.744\pm0.007$   &  $19.714\pm0.012$ \\
A\_17	&   0:15:51.42	&   -39:15:57.60	&   B1I		&   79	&   $20.426\pm0.007$   &  $20.157\pm0.008$ \\
D\_27	&   0:14:28.73	&   -39:10:17.76	&   B2I		&   76	&   $19.395\pm0.004$   &  $19.373\pm0.007$ \\
A\_27	&   0:15:58.03	&   -39:14:27.60	&   B2I		&   66	&   $20.140\pm0.006$   &  $20.050\pm0.010$ \\
C1\_53	&   0:15:14.64	&   -39:13:36.12	&   B2.5I	&   45	&   $20.174\pm0.005$   &  $19.837\pm0.007$ \\
B\_31	&   0:15:32.22	&   -39:14:40.20	&   B2.5I	&   95	&   $19.555\pm0.005$   &  $19.561\pm0.007$ \\
A\_26	&   0:15:55.45	&   -39:15:30.60	&   B2.5I	&   96	&   $19.584\pm0.005$   &  $19.197\pm0.005$ \\
A\_11	&   0:15:45.17	&   -39:15:56.15	&   B5I		&   130	&   $19.005\pm0.004$   &  $18.874\pm0.005$ \\
\hline
\end{tabular}
}
\tablefoot{The columns list: (1) star identification; (2) right ascension (hh:mm:ss); (3) declination (dd:mm:ss);  (4) spectral type;  (5) signal-to-noise ratio of the spectrum; (6) apparent V magnitude; (7) apparent I magnitude. New calibrations have reveled  issues in the published NGC~55 photometry catalog in C08, here a revised photometry is shown for the stars targeted (Castro et al. in prep).}
\label{catalog1}
\end{table*}


\subsection{Stellar parameters}

\begin{figure*}[!]
   		\resizebox{\hsize}{!}{\includegraphics[angle=0,width=\textwidth]{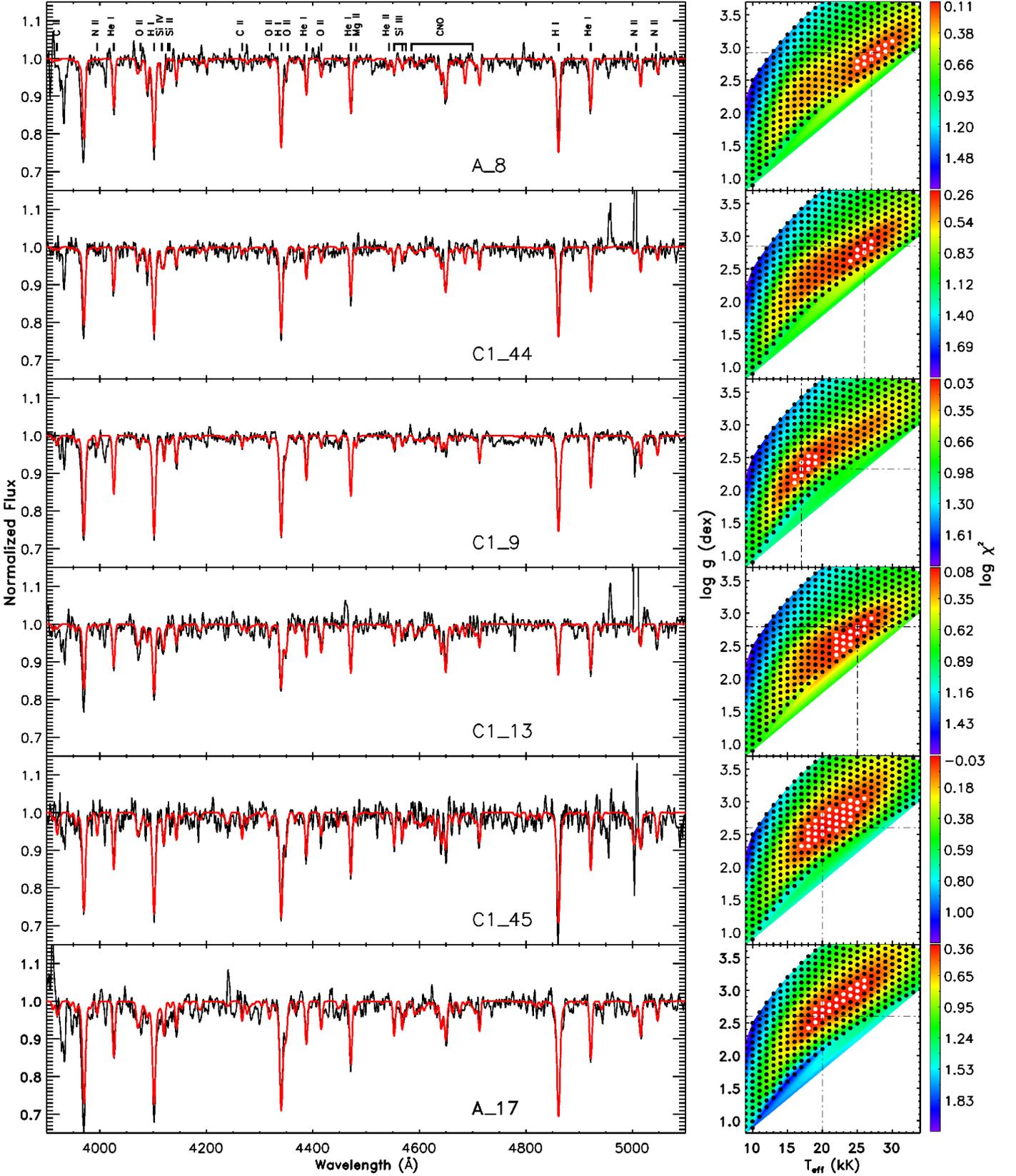}}

   		\caption{Left panels: NGC~55 spectra (black) and best fit model (red) obtained by our automatic method plus the detailed abundance analysis. The lines used in the stellar parameter and chemical abundance determination are marked in the top-left chart. Right panel: $\log\,\chi^2$ difference between observations and the entire grid models in the  $T_{\rm{eff}}-\log\,g$ parameter space. The $\chi^2$ difference is quantified with the color, according to the scale shown at the right of each plot. The black dashed lines mark the model with minimum $\chi^2$. Black dots represent all the grid models; the models whose parameters have been averaged to produce the stellar parameters are highlighted with white dots.}
    \label{Fig:NGC55_stars_page2}
 	\end{figure*}

\begin{figure*}[!]
   		\resizebox{\hsize}{!}{\includegraphics[angle=0,width=\textwidth]{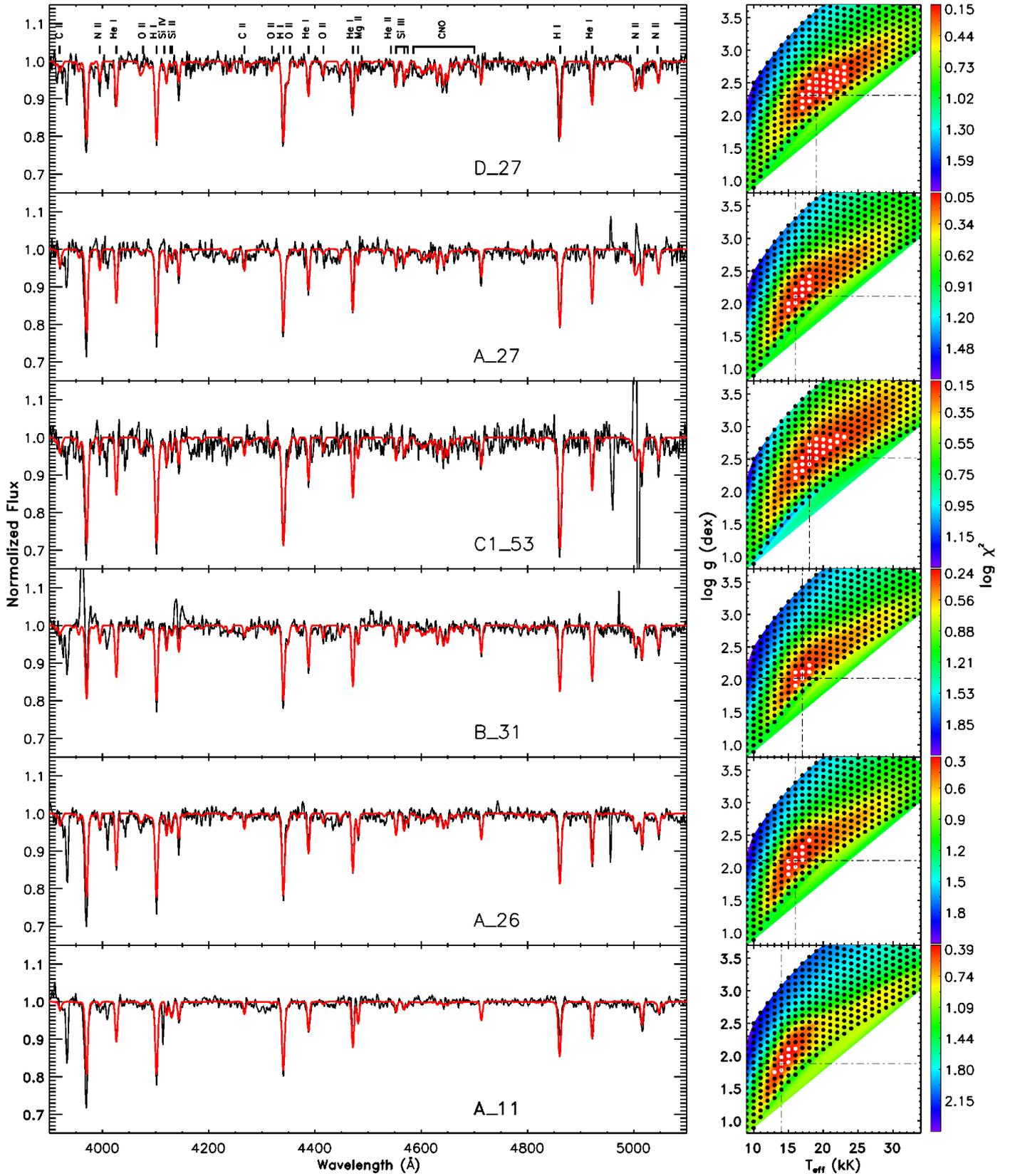}}
   		\caption{Figure \ref{Fig:NGC55_stars_page2}, continued.}
    \label{Fig:NGC55_stars_page3}
 	\end{figure*}

The stellar parameters of our sample of 12 NGC~55 B-type supergiants, obtained from the application of the analysis methodology described in the previous sections, are listed in Table \ref{StPa}. The left side of Figs. \ref{Fig:NGC55_stars_page2} and \ref{Fig:NGC55_stars_page3} show the comparison of the observations with tailored models computed for the parameters derived in this work. It can be seen that these final models provide a good match to the (in some cases rather noisy) observed spectra. The right side of the figures display $\log\,\chi^2$ isocontours on the $T_{\rm{eff}}-\log\,g$ plane. Each black dot represents a model in the grid, whilst the white dots identify those models fulfilling the $\chi^2 - \chi^2_{\mathrm{min}} $ criteria defined in Sect. \ref{metodo_auto}.

The impact that the SNR has on the derived parameters (uncertainties) can be gauged from these plots. For stars like A\_11 and B\_31, the models that reproduce the observations enclose a relatively smaller area in the $T_{\rm{eff}}-\log\,g$ plane than in the case of C1\_45. The higher quality is clearly reflected in the range of stellar parameters, i.e. synthetic models, that are compatible with the observed spectra. Note how the wide range of temperatures and gravities covered by our grid prevented, for our 12 stars, the presence of border-effects, i.e. objects located too close to the limits of the model grid. 

\begin{table*}
\caption{Stellar parameters obtained for the NGC~55 stars by the automatic algorithm presented in this work.  } 
\centering
{\footnotesize
\begin{tabular}{l r r r r r}
\hline\hline
 ID  &   $T_{\rm{eff}}$  &  $\log\,g$ &  $\xi$ & \ion{He}{}/\ion{H}{} & $\log\,Q$  \\  
 (1) &  (2) &   (3)  &   (4)  &   (5)  &  (6)    \\
\hline
A\_8    &	$27700 \pm 1200$	&	$2.91 \pm 0.13$	&	$20.3 \pm 5.8$	&	$0.14 \pm 0.05$	&    $-12.79 \pm 0.33$ \\
C1\_44  &	$26200 \pm 1500$	&	$2.83 \pm 0.15$	&	$17.8 \pm 6.3$	&	$0.10 \pm 0.05$	&    $-12.93 \pm 0.30$ \\
C1\_9   &	$17400 \pm 1500$	&	$2.37 \pm 0.18$	&	$11.8 \pm 3.3$	&	$0.15 \pm 0.05$	&    $-13.37 \pm 0.27$ \\
C1\_13  &	$24200 \pm 1400$	&	$2.69 \pm 0.15$	&	$17.7 \pm 5.7$	&	$0.15 \pm 0.05$	&    $-12.60 \pm 0.20$ \\
C1\_45  &	$21500 \pm 2500$	&	$2.73 \pm 0.24$	&	$19.0 \pm 6.1$	&	$0.12 \pm 0.05$	&    $-13.65 \pm 0.36$ \\
A\_17   &	$22500 \pm 2300$	&	$2.85 \pm 0.16$	&	$20.7 \pm 5.3$	&	$0.10 \pm 0.04$	&    $-13.90 \pm 0.33$ \\
D\_27   &	$19400 \pm 1600$	&	$2.41 \pm 0.15$	&	$14.4 \pm 4.6$	&	$0.10 \pm 0.05$	&    $-13.18 \pm 0.20$ \\
A\_27   &	$16300 \pm 1100$	&	$2.13 \pm 0.19$	&	$20.4 \pm 5.7$	&	$0.12 \pm 0.05$	&    $-12.90 \pm 0.20$ \\
C1\_53  &	$18200 \pm 2000$	&	$2.54 \pm 0.26$	&	$16.4 \pm 5.8$	&	$0.10 \pm 0.05$	&    $-13.73 \pm 0.35$ \\
B\_31   &	$16700 \pm 900$		&	$2.09 \pm 0.13$	&	$20.4 \pm 5.6$	&	$0.14 \pm 0.05$	&    $-12.83 \pm 0.20$ \\
A\_26   &	$16100 \pm 1000$	&	$2.15 \pm 0.17$	&	$17.1 \pm 6.5$	&	$0.12 \pm 0.05$	&    $-12.86 \pm 0.20$ \\
A\_11   &	$14400 \pm 1000$	&	$1.94 \pm 0.18$	&	$11.8 \pm 3.7$	&	$0.10 \pm 0.05$	&    $-12.64 \pm 0.20$ \\
\hline
\end{tabular}}
\tablefoot{ The columns list: (1) star identification; (2) effective temperature (K); (3) surface gravity (dex); (4) microturbulence (km\,s$^-1$); (5) helium abundance (by number); and (6) the Q parameter. } 
\label{StPa}
\end{table*}


\subsection{Chemical abundances}

With the stellar parameters in hand, we performed the analysis of the elemental chemical abundances as described in Sect. \ref{AbunDeter}. The results are compiled in Table \ref{Abun}. The characteristic metallicity, given in the last column, is obtained averaging the differences of the derived O, Mg and Si abundances relative to the solar reference, taken from \citet{2009ARA&A..47..481A}. Projected distances to the galactic center are given in the second column, in units of the semi-major axis.

The final models presented in Figures \ref{Fig:NGC55_stars_page2} and \ref{Fig:NGC55_stars_page3} include the derived elemental abundances. Overall, these figures show a good agreement between the final tailored models, computed for the final abundances,  and the observed spectra. The average abundance uncertainty in these chemical analyses is $\sim$0.25\,dex, with the case of C1\_45, with its poor SNR, reaching 0.30\,dex. Our results indicate that it is possible to chemically characterize these stars with good precision, even for cases of relatively low SNR, such as C1\_53 or C1\_45.

We note here that the derived Mg abundances, based solely on the \ion{Mg}{II} 4481 \AA~feature, could be affected by a blend with \ion{Al}{III} 4479\,\AA. Following the arguments in the works by \cite{2005MNRAS.358..193L} and \cite{2011arXiv1109.6661D}, a small effect would be expected, given the low metallicity of NGC\,55. Moreover, \cite{2005ApJ...635..311U} suggested that there is a negative luminosity dependence of the strength of the \ion{Al}{III} $4479\,\AA$ line, making its contribution even less relevant for supergiant stars, even more since \ion{Mg}{II} $4481\,\AA$ presents the opposite behavior, with the feature strengthening with increasing luminosity.

\begin{table}

\caption{Stellar abundances determined for \ion{Si}{}, \ion{Mg}{}, \ion{C}{}, \ion{N}{} and \ion{O}{}. } 
\begin{center}
{\footnotesize
\begin{tabular}{ l  r  r  r  r  r  r  r}
\hline\hline

ID	&	$\rho/\rho_{o} $	&$\epsilon_{\ion{Si}{}}$\tablefootmark{a}&	$\epsilon_{\ion{C}{}}$\tablefootmark{a}&	$\epsilon_{\ion{N}{}}$\tablefootmark{a}&	$\epsilon_{\ion{O}{}}$\tablefootmark{a}&	$\epsilon_{\ion{Mg}{}}$\tablefootmark{a}	& [\ion{Z}{}/\ion{Z}{}$_\odot$] \\
\hline
$\odot$ &		   & $  7.51 $      &  $ 8.43 $       &  $ 7.83 $       &  $ 8.69 $       &  $7.60 $        &            \\
\hline
A\_8  	&  $ 0.64 $ & $ 7.18 $     & $ 7.27 $	   & $ 7.33  $     & $ 8.28 $	  & $--    $	     &$ -0.37  $ \\
C1\_44  &  $ 0.22 $ & $ 7.08 $     & $ 7.71 $	   & $ 7.56  $     & $ 8.07 $	  & $--    $	     &$ -0.53  $ \\
C1\_9   &  $-0.11 $ & $ 6.80 $     & $ 7.23 $	   & $ 7.95  $     & $ 8.66 $	  & $6.80  $	     &$  -0.51 $  \\
C1\_13  &  $-0.06 $ & $ 7.07 $     & $ 7.60 $	   & $ 7.63  $     & $ 8.45 $	  & $6.92  $	     &$  -0.45 $  \\
C1\_45  &  $ 0.22 $ & $ 7.13 $     & $ 8.07 $	   & $ 8.17  $     & $ 8.34 $	  & $7.00  $	     &$  -0.44 $  \\
A\_17   &  $ 0.75 $ & $ 7.06 $     & $ 7.77 $	   & $ 7.64  $     & $ 8.37 $	  & $7.02  $	     &$  -0.45 $  \\
D\_27   &  $-0.32 $ & $ 7.01 $     & $ 7.37 $	   & $ 8.28  $     & $ 8.26 $	  & $7.27  $	     &$  -0.42 $  \\
A\_27   &  $ 0.80 $ & $ 7.04 $     & $ 7.77 $	   & $ 8.52  $     & $ 8.07 $	  & $6.98  $	     &$  -0.57 $  \\
C1\_53  &  $ 0.28 $ & $ 7.25 $     & $ 7.52 $	   & $ 8.22  $     & $ 8.63 $	  & $7.28  $	     &$  -0.21 $  \\
B\_31   &  $ 0.50 $ & $ 7.10 $     & $ 7.22 $	   & $ 8.25  $     & $ 8.59 $	  & $7.26  $	     &$  -0.28 $  \\ 
A\_26   &  $ 0.78 $ & $ 7.43 $     & $ 7.48 $	   & $ 8.22  $     & $ 8.60 $	  & $7.33  $	     &$  -0.15 $  \\
A\_11   &  $ 0.68 $ & $ 7.17 $     & $ --   $	   & $ --    $     & $ --   $	  & $7.20  $	     &$ -0.37  $ \\

\hline
\end{tabular}
}
\end{center}
\tablefoot{The projected radial distance of the stars to the optical center of NGC~55 normalized by the major semi-axis ($\rho_{o}\sim 16'$) is displayed as well ($\rho/\rho_{o}$). Negative values of $\rho/\rho_{o}$ mean that these stars are located westward of the galactic center. The solar abundances used as a references ($\odot$) were taken from \cite{2009ARA&A..47..481A}.}
\tablefoottext{a}{$\epsilon_{\ion{X}{}}=\log\,(\ion{X}{}/\ion{H}{})+12$ (by number)}
\label{Abun}
\end{table}


\subsection{Comments on individual targets}

Here we discuss some details of individual targets, along with problems encountered during their analysis.

\begin{itemize}

	\item{\textbf{A\_8.} This is an O9.7 I star whose main features are well represented by our final model. At this hot temperature ($27700\,$K) the \ion{Mg}{II} line has vanished and the  \ion{Si}{III} lines are weak. At the same time, \ion{He}{II} lines are present, allowing for a precise determination of the effective temperature. The  \ion{Si}{IV} line, blended with \ion{O}{II}, is well reproduced. The apparent slight discrepancy around the \ion{O}{II} $4079\,\AA$ line is most likely due to an effect of the normalization inside the H$\delta$ wing. Note also that the \ion{O}{II} $4414-16\,\AA$ blend is weaker than the prediction. \ion{He}{II} $4686\ \AA$ shows also a clear mismatch but, without a reliable wind estimation we cannot go further in its analysis. }

	\item{\textbf{C1\_44.} The main spectral features are well reproduced by the final model, with the exception of some transitions like: \ion{He}{II} $4686\ \AA$, possibly affected by the wind, and not considered in the analysis; or \ion{Si}{IV} $4089\ \AA$, not included because of its blend with \ion{O}{II}.  The \ion{O}{II} doublet at $4414-16\ \AA$ is also poorly reproduced, though the rest of the oxygen lines are consistent with the obtained abundance. At this temperature  the \ion{Mg}{II} line is quite weak. The transition of \ion{N}{II} $5007\ \AA$ is filled by nebular contamination.}

	\item{\textbf{C1\_9.} The final model shows a good global match to the majority of the considered transitions. The lines of O, Mg and Si are not particularly strong at this temperature, but they are well reproduced. The temperature derived corresponds to a spectral type cooler than the one assigned.}

	\item{\textbf{C1\_13.} This B1 supergiant presents a very nice match between the final model and the observed spectrum. The blends of \ion{O}{II} $4079\ \AA$ and $4319\ \AA$ show the largest difference, but still within the abundance uncertainties of $\pm$0.25\,dex. Some contamination by nebular emission is still apparent in the  \ion{N}{II} $5007\ \AA$ line. The \ion{Mg}{II} is weak, as expected for this temperature.}

	\item{\textbf{C1\_45.} This spectrum has the lowest SNR ratio in the sample and this is clearly reflected in the  errors derived for the stellar parameters. Figure \ref{Fig:NGC55_stars_page2} shows a large spread of compatible models in the $T_{\rm{eff}}-\log\,g$ plane. Nonetheless, the model corresponding to our solution  reproduces well the main features. The observed \ion{N}{II} $3995\ \AA$ line is not well represented by the model, and  the transition at $5007\ \AA$ is contaminated by nebular emission. As typical for a B1 I,  the \ion{Mg}{II} transition is weak.}

	\item{\textbf{A\_17.} This object shows a discrepancy at the continuum around H$\delta$, likely due to the rectification around H$\delta$ wings. The parameters and abundances derived reproduce well the rest of the spectrum. The magnesium transition is weak. Note in Fig. \ref{Fig:NGC55_stars_page2} the lack of symmetry in $\chi^2$  distribution around the average values. There is a plume of suitable models towards high temperatures that also fulfill our goodness-of-fit criteria.}
	
	\item{\textbf{D\_27.}  The line of \ion{He}{I} at $4144\ \AA$ presents a mismatch. Note that this line is not included in the analysis.}
	
	\item{\textbf{A\_27.} This B2 I  is well represented by the final model. Besides from the mismatch around the $5007\ \AA$ region,  the rest of the spectral features are well reproduced.} 
	
	\item{\textbf{C1\_53.} In spite of the low SNR spectrum obtained for this star, the match between the main features analyzed and the best model is quite good. The spectrum also displays some residuals of nebular contamination at [\ion{O}{III}] $5007\ \AA$.}
	
	\item{\textbf{B\_31.} This B2.5 supergiant star is nicely reproduced by our model, as it is shown in  Fig. \ref{Fig:NGC55_stars_page3}. The main transitions used in the analysis are well reproduced.}

	\item{\textbf{A\_26.} This B2.5 I star shows a good agreement with the model. Its better SNR ratio results in a well constrained set of parameters. The oxygen lines are weaker at this temperature but  they are all well modeled except for \ion{O}{II} $4079\ \AA$. The latter issue may be caused by the continuum rectification, as in A\_17, or residuals from cosmic ray subtraction.}

	\item{\textbf{A\_11.} This star has the best SNR spectrum, and this is reflected in the small uncertainties. Its effective temperature suggests a spectral type a bit earlier than the one proposed by C08. At these cool temperatures the transitions of \ion{O}{II}, \ion{N}{II} and \ion{C}{II} are very weak or have even completely vanished. The \ion{Si}{III} lines  are weak but the transitions of \ion{Si}{II} (although blended with \ion{O}{II} lines) become stronger and allow us to derive effective temperature and constrain the silicon abundance.}

\end{itemize}  

\section{Discussion}
\label{metallicity}

In this section we discuss the results obtained for our sample of 12 B-type supergiant stars in the galaxy NGC~55. The physical characterization provided by the stellar parameters and chemical abundances, supplies us with the necessary information to discuss their evolutionary status, as well as to carry out a comparison with the predictions of current evolutionary models. We adopt a distance modulus to NGC~55 of $\mu=26.434\pm0.037$ mag from \citet{2008ApJ...672..266G}. 


\subsection{Stellar properties}
\label{Stellar_properties}

\begin{table*}
\caption{Photometry and fundamental parameters recovered from the sample analyzed in NGC~55. } 

\centering
{\footnotesize
\begin{tabular}{l r r r r r r r r}
\hline\hline
 ID  &   $\log\,g_F$  &  $M_\mathrm{bol}$ &  $E(B-V)$ & $BC$ & $\log\,L/L_\odot$ & $R/R_\odot$  & $M_\mathrm{Spec}/M_\odot$ & $M_\mathrm{Evol}/M_\odot$ \\  

 (1) &  (2) &   (3)  &   (4)  &   (5)  &  (6) & (7) & (8) & (9)   \\

\hline

A\_8	&$ 1.14  \pm 0.07$&$   -9.60 \pm 0.11 $&$   0.177 \pm 0.014 $&$ 	-2.686 \pm  0.096$&$    5.74 \pm 0.04    $&$   32.2 \pm  1.1   $&$    30.7 \pm 10.3   $&$   32.7 \pm 3.5 $\\
C1\_44	&$ 1.16  \pm 0.07$&$  -10.24 \pm 0.11 $&$   0.199 \pm 0.017 $&$ 	-2.547 \pm  0.087$&$    5.99 \pm 0.04    $&$   48.2 \pm  3.1   $&$    57.4 \pm 23.5   $&$   46.8 \pm 5.8 $\\
C1\_9	&$ 1.41  \pm 0.07$&$   -8.57 \pm 0.13 $&$  -0.049 \pm 0.016 $&$ 	-1.592 \pm  0.119$&$    5.32 \pm 0.05    $&$   50.7 \pm  5.6   $&$    22.0 \pm 11.6   $&$   21.7 \pm 1.8 $\\
C1\_13	&$ 1.15  \pm 0.09$&$   -9.98 \pm 0.15 $&$   0.126 \pm 0.020 $&$ 	-2.375 \pm  0.130$&$    5.89 \pm 0.06    $&$   50.1 \pm  2.4   $&$    44.9 \pm 17.6   $&$   40.1 \pm 4.7 $\\
C1\_45	&$ 1.40  \pm 0.12$&$   -9.26 \pm 0.24 $&$   0.171 \pm 0.024 $&$ 	-2.041 \pm  0.222$&$    5.60 \pm 0.09    $&$   45.6 \pm  5.7   $&$    40.9 \pm 27.7   $&$   28.5 \pm 4.0 $\\
A\_17	&$ 1.44  \pm 0.08$&$   -9.36 \pm 0.26 $&$   0.377 \pm 0.026 $&$ 	-2.185 \pm  0.249$&$    5.64 \pm 0.11    $&$   43.6 \pm  3.7   $&$    49.3 \pm 22.3   $&$   29.8 \pm 4.3 $\\
D\_27	&$ 1.26  \pm 0.10$&$   -9.29 \pm 0.22 $&$   0.137 \pm 0.021 $&$ 	-1.824 \pm  0.206$&$    5.61 \pm 0.09    $&$   56.7 \pm  3.7   $&$    30.2 \pm 12.4   $&$   28.7 \pm 3.5 $\\
A\_27	&$ 1.28  \pm 0.09$&$   -8.16 \pm 0.13 $&$   0.151 \pm 0.019 $&$ 	-1.393 \pm  0.114$&$    5.16 \pm 0.05    $&$   47.7 \pm  3.5   $&$    11.2 \pm  5.7   $&$   18.7 \pm 1.6 $\\
C1\_53	&$ 1.50  \pm 0.11$&$   -9.13 \pm 0.21 $&$   0.394 \pm 0.021 $&$ 	-1.653 \pm  0.195$&$    5.55 \pm 0.08    $&$   60.1 \pm  7.5   $&$    45.7 \pm 33.1   $&$   26.0 \pm 3.0 $\\
B\_31	&$ 1.20  \pm 0.08$&$   -8.58 \pm 0.12 $&$   0.078 \pm 0.018 $&$ 	-1.460 \pm  0.096$&$    5.33 \pm 0.05    $&$   55.3 \pm  3.0   $&$    13.7 \pm  4.9   $&$   21.1 \pm 1.6 $\\
A\_26	&$ 1.32  \pm 0.09$&$   -9.44 \pm 0.13 $&$   0.392 \pm 0.017 $&$ 	-1.370 \pm  0.111$&$    5.67 \pm 0.05    $&$   88.2 \pm  5.8   $&$    40.1 \pm 18.3   $&$   29.2 \pm 3.0 $\\
A\_11	&$ 1.31  \pm 0.08$&$   -9.07 \pm 0.12 $&$   0.153 \pm 0.014 $&$ 	-1.164 \pm  0.111$&$    5.52 \pm 0.05    $&$   93.0 \pm  7.6   $&$    27.5 \pm 13.7   $&$   25.9 \pm 2.2 $\\

\hline
\end{tabular}}
\tablefoot{The columns list: (1) star identification; (2) flux-weighted gravity (dex); (3) bolometric magnitude; (4) color excess in (B-V);  (5) bolometric correction;  (6) stellar luminosity; (7) radius; (8) spectroscopic mass; and (9) evolutionary mass.}

\label{Photo}
\end{table*}

Table \ref{Photo} gathers the fundamental stellar properties derived for the stars. The color excess of each object was calculated using observed photometry (see Table \ref{catalog1}) and the synthetic colors obtained from the final tailored models, adopting the extinction curve by \cite{1989ApJ...345..245C} and a total-to-selective-extinction ratio R$_\mathrm{v}=3.1$, although several authors have shown that high $R_v$ are not rare for massive stars (see, for instance, \citealt{2011ApJ...729L...9B} or \citealt{2011A&A...530L..14B}). Only for one of the objects, C1\_9, we find a non-physical value (i.e. negative), although very small and compatible with zero within the error bars. The high inclination of the galaxy, as well as the fact that these objects could be, to some extent, surrounded by ionized gas, could bias the observed photometry which would explain this negative value, an effect also suggested by \cite{2007ApJ...659.1198E}. Three other objects, A\_17, A\_26 and C\_53, present high reddening values, $\sim$0.4 mag, whilst the rest of the sample show $E(B-V)$ values consistent with the mean value derived by  \cite{2008ApJ...672..266G} , $E(B-V)=0.127\pm0.019$ mag, from multi-wavelength observations of Cepheid stars.

\citet{1983MNRAS.204..743W} studied a sample of 7 \ion{H}{II} regions distributed along the southern half of NGC~55, approximately covering the same range in galactocentric distances as our stellar sample. From their published $C\left(H\beta\right)$ values we can obtain reddening values by adopting $E(B-V) = 0.676\,C\left(H\beta\right)$ \citep{1979MNRAS.187P..73S}. The \ion{H}{II} regions present reddening in the range 0.16\,--\,0.32 mag, with a simple mean of $E(B-V)=0.24\pm0.07$ mag, with 3 \ion{H}{II} regions showing $E(B-V)>0.3$ mag. The mean value of our sample, $E(B-V)=0.20\pm0.14$ mag, compares well with the nebular mean, albeit with a larger scatter.

	\begin{figure}[!]
   		\resizebox{\hsize}{!}{\includegraphics[angle=0,width=\textwidth]{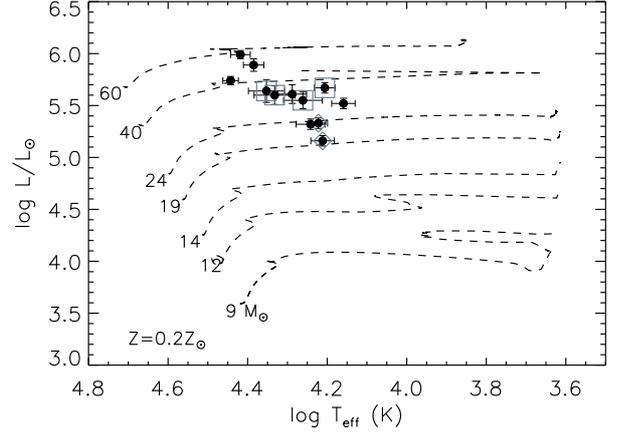}}
   		\caption{The sample stars in the HR diagram. Dashed lines mark the evolutionary tracks of \cite{2001A&A...373..555M} at SMC metallicity with an initial rotational velocity of $300\,\rm{km\,s^{-1}}$. Those stars that have shown large discrepancies between the spectroscopic mass and the evolutionary one are marked by gray squares or diamonds (see text for additional details).}
    	\label{Fig:NGC55_HR}
 	\end{figure}


	\begin{figure*}
   		\resizebox{\hsize}{!}{\includegraphics[angle=90,width=\textwidth]{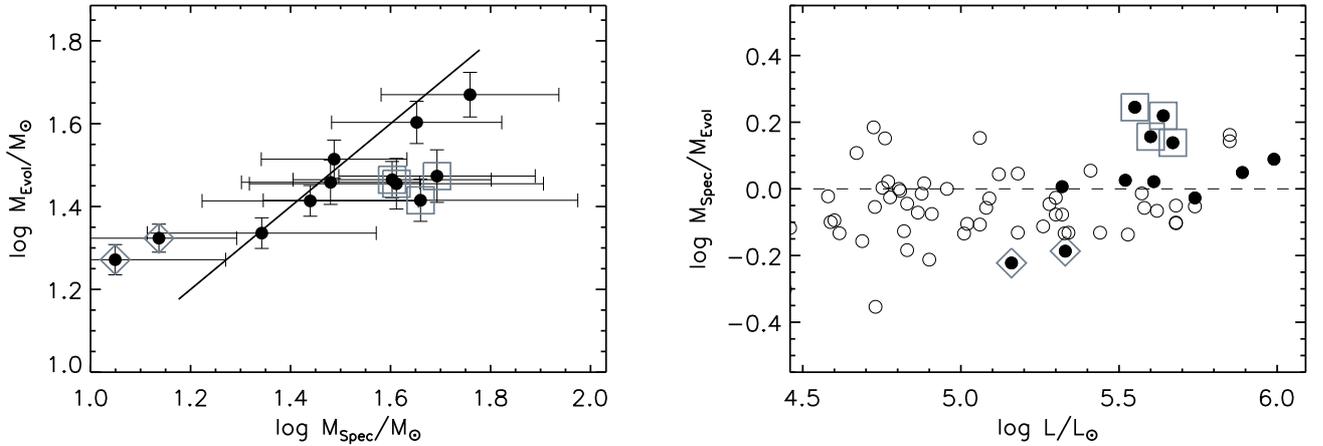}}
   		\caption{Left: Relationship between the spectroscopic and evolutionary masses derived in NGC~55 (the one-to-one trend is marked with a black solid line). Right: The logarithm ratio of masses versus their luminosity in this work (black dots), together with the results published previously in other galaxies by \cite{2008ApJ...681..269K}, \cite{2008ApJ...684..118U} and \cite{2009ApJ...704.1120U} (open circles). Those objects that show higher evolutionary mass compared with the spectroscopic counterpart are highlighted with gray squares for quick contrast with Fig. \ref{Fig:NGC55_HR}. The opposite behavior is indicated with gray diamonds.}
    	\label{Fig:MM}
 	\end{figure*}

The distribution of the stars in the HR diagram is shown in Fig. \ref{Fig:NGC55_HR}, together with the evolutionary tracks by \cite{2001A&A...373..555M}. According to the metallicity of the sample (Table \ref{Abun}), we consider a linear interpolation between  evolutionary tracks computed for the Small Magellanic Cloud metallicity \citep{2001A&A...373..555M} and solar metallicity tracks \citep{2005A&A...429..581M}. The evolutionary masses derived from the interpolation of these tracks are shown in the last column of Table \ref{Photo}. These results were also checked with the recent evolutionary calculations for LMC metallicity by \cite{2011A&A...530A.115B}, finding very similar results.

The comparison of spectroscopic and evolutionary masses has been an important source of discrepancy between the evolutionary theories and the stellar atmosphere modeling for decades \citep{1992A&A...261..209H}. Improvements in both fields have minimized this issue. Nonetheless, a systematic shift can still be measured in the analysis of B-type supergiant stars. For instance, the analysis carried out by \cite{2009ApJ...704.1120U} in M~33 revealed an average difference of $0.06\,$dex between spectroscopic and evolutionary masses. Although small, it is still a systematic issue. Figure \ref{Fig:MM} displays the relationship between these two mass estimations for our NGC~55 B-type supergiant stars. The errors in the spectroscopic masses are large enough to make the full sample consistent with the one-to-one relation (left panel in Fig. \ref{Fig:MM}); half of the sample shows a very good agreement between both measurements and no systematic trend is evident, whilst four stars (A\_17, A\_26, C1\_45 and C1\_53, marked in the figure with gray squares) show significantly higher evolutionary than spectroscopic masses. Table \ref{Photo}  reveals that three of them show the highest color excess in the sample, which could point to a nebular contamination effect in the observed photometry. On the other hand, A\_27 and B\_31 show the opposite behavior (marked with gray diamonds in Fig. \ref{Fig:MM}). Due to the spatial and spectroscopic resolution we cannot rule out additional unresolved companion(s) that would be affecting the photometry (but are not evident in the optical spectra).


\subsection{Flux-weighted Gravity--Luminosity Relationship}

\begin{figure}
   		\resizebox{\hsize}{!}{\includegraphics[angle=0,width=\textwidth]{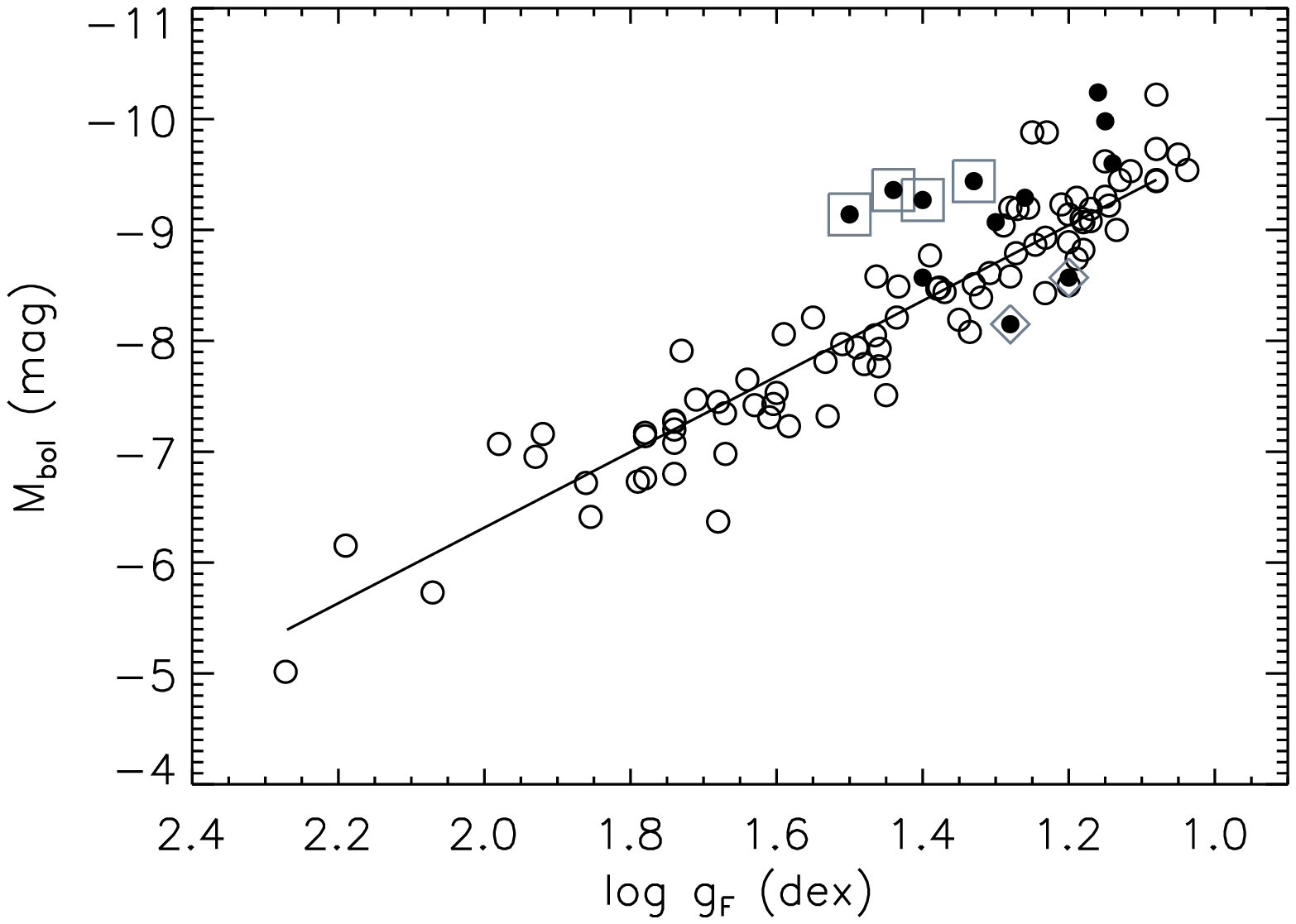}}
   		\caption{Position for our NGC~55 sample in the FGLR (black dots). Previous studies performed by \cite{2008ApJ...681..269K}, \cite{2008ApJ...684..118U} and \cite{2009ApJ...704.1120U} in different galaxies are also displayed (open circles), the linear relationship proposed by  \cite{2008ApJ...681..269K} (see Eq. \ref{Eq:log_gf}) is drawn with a solid black line. The NGC~55 stars that have shown stronger discrepancies between the spectroscopic and evolutionary masses are marked as in Fig. \ref{Fig:NGC55_HR} (see Sect. \ref{Stellar_properties} for additional details). }
    	\label{Fig:FGLR}
 	\end{figure}

Figure \ref{Fig:FGLR} shows the flux weighted gravity--luminosity relationship for the NGC~55 stars, together with the results published by \cite{2008ApJ...681..269K}, \cite{2008ApJ...684..118U} and \cite{2009ApJ...704.1120U}. The location of the NGC~55 stars shows a good agreement with these studies, worst for those stars for which we have found disagreements between the different mass estimations. The other six stars follow the same trend with a slight shift towards higher bolometric magnitudes, although they are within the observed scatter of the distribution \citep{2008ApJ...681..269K}. An independent determination of the distance to NGC~55 based on the FGLR is deferred to a future publication, since information from BA supergiant stars, with $\log\,g_F>1.5$ dex, is mandatory to properly determine the distance modulus.


\subsection{Evolutionary chemical status}

	\begin{figure*}
   		\resizebox{\hsize}{!}{\includegraphics[angle=90,width=\textwidth]{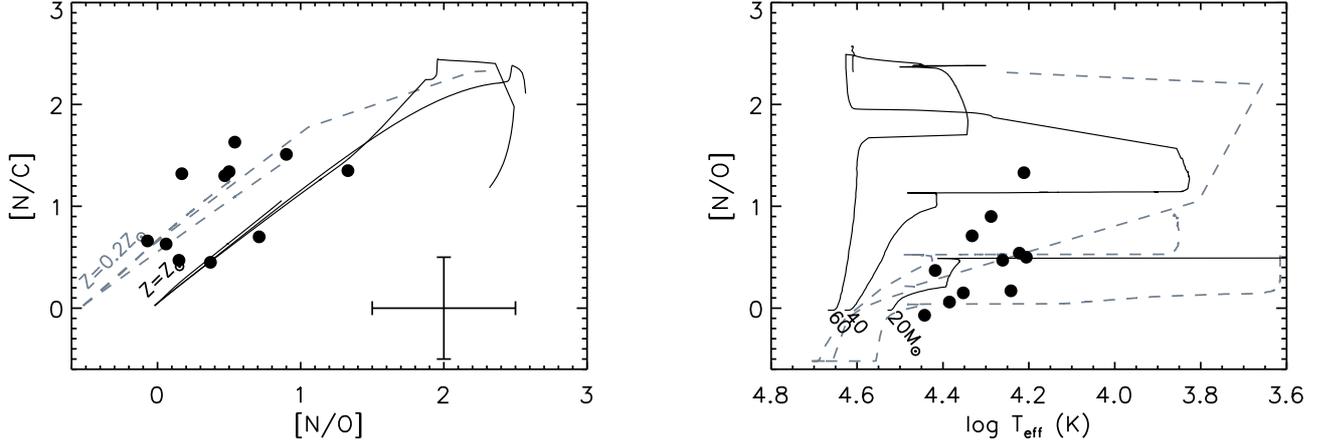}}
   		\caption{Left panel: [\ion{N}{}/\ion{C}{}] versus [\ion{N}{}/\ion{O}{}] ratios (by number) relative to solar values derived in our NGC~55 sample, the average uncertainty is plotted  in the right-bottom corner. The evolutionary tracks for three different masses ($20$, $40$ and $60\,M_\odot$), are displayed for solar (solid black lines, \citealt{2005A&A...429..581M}) and  SMC metallicity (dashed gray lines, \citealt{2001A&A...373..555M}). Right panel: position of the sample stars in the [\ion{N}{}/\ion{O}{}]\,--\,$T_{\rm{eff}}$ diagram, using the same color and symbol code that in the left panel.}
    	\label{Fig:NGC55_CNO_evol}
 	\end{figure*}

Current massive star evolutionary models, accounting for the effects of mass-loss and rotation, predict a tight relationship between \ion{N}{}/\ion{C}{} and \ion{N}{}/\ion{O}{} surface ratios as a consequence of the mixing with CNO processed material from inner layers during the stellar evolution (e.g. \citealt{1986ARA&A..24..329C}). Previous studies have observationally found this relationship in Galactic stars (see for instance \citealt{2010A&A...517A..38P} and references therein), although the large uncertainties leave open a broad range of interpretations. The detailed study presented by \cite{2010A&A...517A..38P} on a sample of Galactic B-type dwarfs and A-type supergiant stars of the solar neighborhood revealed a very good agreement with the theoretical predictions of single star evolutionary models, in the particular range of stellar masses sampled by these objects, $\sim$20\,--\,40\,$M_\odot$. It has to be still proven that this is also the case for other mass ranges. Our sample in NGC~55 contains an heterogeneous group of stars throughout the disc of NGC~55, nonetheless the left panel of Fig. \ref{Fig:NGC55_CNO_evol} shows that these stars follow the theoretical predictions in a qualitative sense. 

It would seem that our derived N abundances define two groups of stars, with the first one clustering around 7.62$\pm$0.22 dex (simple mean and standard deviation) and a second one at 8.28$\pm$0.125 dex. Adopting as N baseline the \ion{H}{II} region values derived by \citet{1983MNRAS.204..743W}, 6.63$\pm$0.10 dex and a mean N/O\,=\,0.015, all our stars show a high degree of N processing: N/O\,=\,0.15 and 0.69 (mean values for each group). Interestingly enough, similar values of enrichment have been recently reported for O type stars in the LMC by \cite{2011arXiv1110.5148R}. These authors find two distinct groups of strongly enriched objects, with N abundances 7.5 dex and 8.1 dex, with the LMC baseline N abundance being at 6.9 dex. Our B-type supergiant stars in NGC\,55 show a remarkable agreement with the N abundances of O type stars in the LMC. This strongly supports the idea that our B supergiant stars belong to a young population evolving away from the Main Sequence towards the red part of the HR diagram. We see no indication of any object being in a blue-loop, i.e. being a post Red Supergiant object: besides the previous discussion, current evolutionary models predict that the blue loops cannot reach the temperatures of our objects. Therefore, the main conclusion is then that none of these objects is in an advanced evolutionary stage.


\subsection{Metallicity distribution in the disk of NGC~55}

	\begin{figure*}
   		\resizebox{\hsize}{!}{\includegraphics[angle=0,width=\textwidth]{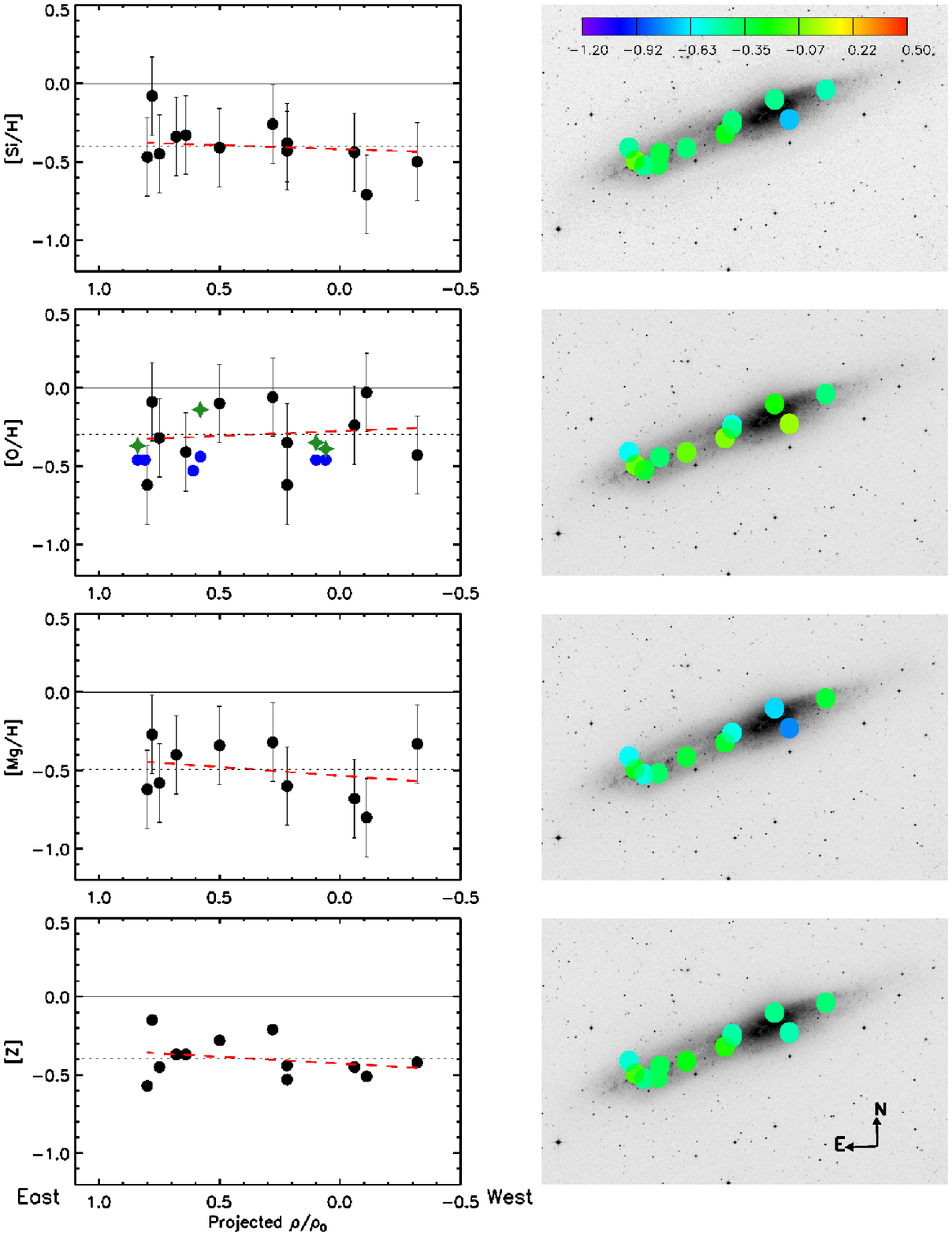}}
   		\caption{Spatial distribution of chemical species in NGC~55. The plots in the left column display the abundance of \ion{Si}{}, \ion{Mg}{}, \ion{O}{} and \ion{Z}{}  with respect to solar values \citep{2009ARA&A..47..481A} against projected galactocentric distance ($\rho/\rho_{o}$). The mean value is indicated with a dotted black line and the linear regression with a red dashed line. The linear regressions do not show any clear abundance pattern with  almost null slopes (see Table \ref{grad}). For oxygen, we also show the \ion{H}{II} regions oxygen abundances (blue dots) derived by \cite{1983MNRAS.204..743W} based on the application of strong-line methods. New estimation of \ion{O}{} abundances, obtained in this work, based on the auroral line [\ion{O}{III}] 4363\,$\AA$ measured by \cite{1983MNRAS.204..743W} (green stars, see text for additional details). The plots in the right column show the 2D distribution maps, following the color-code shown in the top-right plot. }
    	\label{Fig:NGC55_CNO}
 	\end{figure*}

\begin{table}
\caption{Element abundances of \ion{Si}{},  \ion{O}{}, \ion{Mg}{} and  estimated \ion{Z}{} averaged over the whole sample. }
\centering
\begin{tabular}{l  r  r  r}
\hline\hline
[X/H]   &  Mean$\,\pm\,\sigma$   &\textit{a}     &   \textit{b}     \\
\hline
\ion{Si}{}      & $   -0.40\pm0.15 $   &  $   0.05\pm0.07 $   &  $  -0.42\pm0.03 $  \\
\ion{O}{}       & $   -0.30\pm0.21 $   &  $  -0.06\pm0.18 $   &  $  -0.28\pm0.09 $  \\  
\ion{Mg}{}      & $   -0.49\pm0.18 $   &  $   0.11\pm0.15 $   &  $  -0.53\pm0.08 $  \\
\ion{Z}{}       & $   -0.40\pm0.13 $   &  $   0.09\pm0.10 $   &  $  -0.43\pm0.05 $  \\
\hline
\end{tabular}
\tablefoot{The columns $a$ and $b$ provide the coefficients of the linear regression (see Fig. \ref{Fig:NGC55_CNO}), which accounts for the spatial chemical distribution.}
\label{grad}
\end{table}

Previous works based on the study of the emission line spectra of \ion{H}{II} regions have found that the present-day metallicity (oxygen) of NGC~55 is very similar to that of the LMC \citep{1983MNRAS.204..743W,2005ApJ...622..279D}. From our sample of 12 B-type supergiant stars we find a mean metallicity of $-0.40\pm0.13\,$dex  (see the first column in Table \ref{grad}), a value quite close to the LMC metallicity \citep{2007A&A...466..277H}. We reached a similar conclusion with our previous qualitative analysis (C08).

We also calculated the \ion{O}{}/\ion{H}{} abundance ratio for the four \ion{H}{II} regions for which \cite{1983MNRAS.204..743W} reported the detection of the [\ion{O}{III}] 4363\,$\AA$ auroral line. This allows us to provide `direct' nebular abundances, which are independent of various calibrations of strong-line methods. Electron temperatures were calculated with the {\it temden} program in IRAF\footnote{IRAF is distributed by the National Optical Astronomy Observatory, which is operated by the Association of Universities for Research in Astronomy, Inc., under cooperative agreement with  the National Science Foundation.}, using the [\ion{O}{III}] $4363/(4959+5007)$ line ratio and updated atomic parameters, as in \cite{2009ApJ...700..309B}. The \ion{O}{}$^{+}$ and \ion{O}{}$^{++}$ ionic abundances were then calculated with the program {\it ionic}. We then obtained \ion{O}{}/\ion{H}{} as result of the sum of these two ionic abundances. Figure \ref{Fig:NGC55_CNO} displays the excellent agreement between these new abundances (shown as stars in the figure) and the ones derived from our sample of B-type supergiants.

The inclination of NGC~55, along with its apparently irregular shape, make this galaxy a difficult object for morphological classification.  The presence of a metallicity gradient across the galaxy would hint to a possible spiral disc. The study of \cite{1983MNRAS.204..743W} did not find any trace of spatial variations in the southern part of the galaxy. To investigate this issue, the distribution of silicon, oxygen and magnesium were measured from our sample of B-type supergiants. We show the spatial trends in Fig. \ref{Fig:NGC55_CNO}.  The individual elemental abundances hint to an almost null gradient. In order to quantify these results, we fit radially dependent gradients, $ [\ion{X/H}{}]=a\,(\rho/\rho_{o} )+b$, to our stellar data. We exclude from the fits  those objects with values beyond $\pm2\,\sigma$ of the mean value. In the previous expression, $[\ion{X/H}{}]=\log\,(\ion{X/H}{})-\log\,(\ion{X/H}{})_{\odot}$, $X/H$ represents the abundance of each element relative to H by number, $(X/H)_{\odot}$ is the solar reference, and $\rho$ is the projected galactocentric distance (the semi-major axis, $\rho_{o}\sim16'$ for NGC~55). The parameters of the linear regression, $a$ and $b$ are collected in Table \ref{grad}. Considering the errors in the regression coefficients (Table \ref{grad}), all the elements show a spatial distribution consistent with no gradient, supporting the results by \cite{1983MNRAS.204..743W}, but in this case based not only O, but also Mg and Si.  Nonetheless, it would be highly desirable to expand the sample in the western half of the galaxy, before drawing a definitive conclusion. The right side of Fig. \ref{Fig:NGC55_CNO} shows the 2D abundance distribution over NGC~55. There is not a clear abundance pattern when 2 dimensions are considered either. We cannot rule out any projection effects that could blur a chemical gradient due to the high inclination of NGC~55.


\section{Summary and Conclusions}
\label{Conclusions}
 
Motivated by the necessity of analyzing large samples of optical spectra of massive blue stars in an objective way, even for the case of low spectral resolution data, we have undertaken the steps to implement a grid-based methodology. We first computed an extensive model grid with the model atmosphere/line formation code {\sc fastwind}. This new grid was specifically designed for the spectral analysis of blue supergiant stars of spectral types O9 to A0. Secondly, we implemented an algorithm that determines the stellar parameters by finding the subset of models in the grid fulfilling the criteria of minimizing the differences with respect to the observed spectrum.

We have shown, through a number of control tests, that our methodology is well suited for the analysis of optical spectra of B-type supergiants, even in the low spectral resolution case. The analysis of synthetic models (degraded to the expected observational conditions), showed that the stellar parameters are recovered with a high degree of fidelity. Furthermore, the study of three Galactic stars, degraded to low resolution and SNR, provided answers that are consistent with results based on high spectral resolution analysis present in the literature. 
 
As a first application of our model grid and analysis algorithm we have analyzed a sample of 12 early B-type supergiants located in the Sculptor filament galaxy NGC~55. Our methodology allowed us to obtain a complete  characterization of these stars, in terms of their stellar parameters and surface chemical composition, in spite of the low spectral resolution and SNR. The tailored final models provided an accurate match to the observations.

Half of the objects in our sample presented a good agreement between the evolutionary and spectroscopic masses. For the rest, the agreement is not so good, but the results could be considered in agreement within the uncertainties of the analysis. The location of the stars in the FGLR, for the adopted distance to NGC~55, showed a good correspondence with results obtained in previous studies.

The average metallicity of the sample is $\log\,\left(Z/Z_\odot\right)\sim-0.40$ dex. Our results indicate that NGC~55 does not sustain radial abundance gradients, thus confirming previous works based on \ion{H}{II} regions. Nonetheless, the inclination and morphological structure of NGC~55 make this galaxy an interesting target for studies of the 2D metallicity distribution, key for understanding the chemical evolution of NGC~55. The derived CNO compositions show that our stars are evolving away from the Main Sequence, and that none of these objects is returning from an excursion to the red side of the HR diagram. We have found an apparent separation in the nitrogen abundances in two groups, both being strongly enriched in comparison with the N baseline abundance defined by the \ion{H}{II} regions. The derived values are in good agreement with a recent study of LMC O type stars by \cite{2011arXiv1110.5148R}, strongly supporting the idea that our objects are evolving directly from the Main Sequence.

We have shown the reliability of our new automatic, objective and fast methodology for the analysis of massive blue stars, even at low spectral resolution. Its application to large samples will enable us to tackle different issues in a statistical and systematic way. In future work, we will apply this method 
to an extended sample of massive blue stars in NGC~55. This analysis will provide us with additional information on the discrepancy of masses for B-type supergiant stars, their evolution, the FGLR and a detailed description of the 2D galactic chemical distribution.


\begin{acknowledgements}

The authors would like to thank the referee, I. Hunter,  for his useful comments and very helpful suggestions to improve this paper. A. Z. Bonanos is also acknowledged for her careful reading of the manuscript. This project has been supported by Spanish grants number AYA2008-06166-C03-01, AYA2010-21697-C05-04 and was partially funded by the Spanish MICINN under the Consolider-Ingenio 2010 Program grant CSD2006-00070 (http://www.iac.es/consolider-ingenio-gtc) and the Gobierno Aut\'onomo de Canarias under project PID2010119. NC acknowledges research and travel support from the European Commission Framework Program Seven under the Marie Curie International Reintegration Grant PIRG04-GA-2008-239335. WG and GP gratefully acknowledge financial support for this work from the Chilean Center for Astrophysics FONDAP 15010003, and from the BASAL Centro de Astrof\'{\i}sica y Tecnolog\'{\i}as Afines (CATA) PFB-06/2007. MAU, FB and RPK were supported by the National Science Foundation under grant AST-1008798. In addition, RPK acknowledges support by the Alexander-von-Humboldt Foundation and the hospitality of the MPI for Astrophysics and the University Observatory Munich, where part of this work were carried out. The authors would like to thank the Instituto de Astrof\'{i}sica de Canarias computer network and CONDOR (http://www.cs.wisc.edu/condor) facilities. Support from the FOCUS and TEAM subsidies of the Foundation for Polish Science (FNP) and the Ideas Plus grant of the Ministry of Science and Higher Education is also acknowledged.

\end{acknowledgements}


\bibliographystyle{aa}

\bibliography{AA_18253-11_ncastro}

\end{document}